\documentclass[prb,twocolumn,showpacs,amsmath,amssymb,superscriptaddress]{revtex4-1}
\usepackage{graphicx}
\usepackage{dcolumn}
\usepackage{bm}
\usepackage{bbm}
\usepackage{ulem}
\usepackage{color}

\begin{document}

\title{Quantum Anomalous Hall Effect in Magnetically Doped InAs/GaSb Quantum Wells}

\author{Qingze Wang$^1$, Xin Liu$^1$, Hai-Jun Zhang$^2$, Nitin Samarth$^1$, Shou-Cheng Zhang$^2$ and Chao-Xing Liu}
\affiliation{Department of Physics, The Pennsylvania State University, University Park, Pennsylvania 16802-6300\\
$^2$ Department of Physics, McCullough Building, Stanford University, Stanford, CA 94305-4045}

\date{\today}

\begin{abstract}
The quantum anomalous Hall effect has recently been observed experimentally in thin films of Cr doped (Bi,Sb)$_2$Te$_3$ at a low temperature ($\sim$ 30mK). In this work, we propose realizing the quantum anomalous Hall effect in more conventional diluted magnetic semiconductors with doped InAs/GaSb type II quantum wells. Based on a four band model, we find an enhancement of the Curie temperature of ferromagnetism due to band edge singularities in the inverted regime of InAs/GaSb quantum wells. Below the Curie temperature, the quantum anomalous Hall effect is confirmed by the direct calculation of Hall conductance. The parameter regime for the quantum anomalous Hall phase is identified based on the eight-band Kane model.
The high sample quality and strong exchange coupling make magnetically doped InAs/GaSb quantum wells good candidates for realizing the quantum anomalous Hall insulator at a high temperature.
\end{abstract}

\maketitle

{\it Introduction} - The quantum anomalous Hall (QAH) state in magnetic topological insulators\cite{Haldaneprl1988,Qiprb2006,Qiprb2008,Liucprl20082,Wuprl2008,Yusci2010,Dingprb2011,Zhangprl2012,Changsci2013,Wangzfprl2013,Wangjprl2013,Zhangscirep2013} possesses a quantized Hall conductance carried by chiral edge states, similar to the well-known quantum Hall state\cite{Klitzingprl1980}. However, its physical origin is due to the exchange coupling between electron spin and magnetization, instead of the orbital effect of magnetic fields. Therefore, the QAH effect does not require any external magnetic field or the associated Landau levels\cite{Haldaneprl1988}, and thus has great potential in the application of a new generation of electronic devices with low dissipation.

Nevertheless, it is difficult to search for realistic systems of the QAH effect, mainly due to the stringent material requirements. To realize the QAH effect, the system should be an insulator with a topologically non-trivial band structure. Simultaneously, ferromagnetism is also required in the same system. In realistic materials, it is rare that ferromagnetism coexists with an insulating behavior. For example, the HgTe/CdTe quantum well (QW) is the first quantum spin Hall (QSH) insulator with a topologically non-trivial band structure\cite{Bernevigsci2006,Konigsci2007}. With magnetization, it was also predicted to be a QAH insulator\cite{Liucprl20082}. However, ferromagnetism cannot be developed spontaneously in this system. Thus, one has to apply an external magnetic field, obstructing the confirmation of the QAH effect. Alternatively, one can consider other two dimensional (2D) topological insulators with magnetic doping. It was realized that the non-trivial band structure in Bi$_2$Te$_3$ family of materials can significantly enhance spin susceptibility and may lead to ferromagnetism in the insulating state\cite{Yusci2010}. Consequently, these systems with magnetic doping become a suitable platform to observe the QAH effect. The recent experimental discovery\cite{Changsci2013} confirmed this prediction by transport measurements on thin films of Cr doped (Bi,Sb)$_2$Te$_3$. In this experiment, the quantized Hall conductance was observed at a low temperature, around $\sim 30mK$, presumably due to the small band gap opened by exchange coupling and low carrier mobility $\sim 760 cm^2/Vs$ . Therefore, searching for realistic systems with non-trivial band structures, strong exchange coupling and high sample quality is essential to realize the QAH effect at a higher temperature. The QAH effect has also been theoretically predicted in other types of systems\cite{Zhangprl2012,Wuprl2008,Dingprb2011,Zhangscirep2013,Liuxprl2013,Hsuprb2013,Zhangharxiv2013}.

Here we propose a new system for the QAH effect, magnetically doped InAs/GaSb type II QWs. The InAs/GaSb QW is predicted to be a QSH insulator with a certain range of well thickness\cite{Liucprl2008}. Transport experiments have indeed observed a stable longitudinal conductance plateau with a value of $2e^2/h$ in this system\cite{Knezprb2010,Knezprl2011,Suzukiprb2013,Dularxiv2013}. Moreover, Mn-doped InAs and GaSb are known diluted magnetic semiconductors\cite{von1991,Ohnoprl1992,NishitaniPE2010}.  High quality heterostructures integrating (In,Mn)As and GaSb have been fabricated by molecular beam epitaxy\cite{munekata1993}. Unlike the case of Mn-doped GaAs where debate continues about the origin of ferromagnetic ordering \cite{samarth2012,chapler2012,fujii2013}, the ferromagnetism in Mn-doped InAs is consistent with free hole mediated ferromagnetism within a mean field Zener model \cite{Dietlsci2000,macdonald2005,Jungwirthrmp2006,Dietlnm2010}. Magnetically-doped InAs/GaSb heterostructures are also attractive for optical control \cite{koshihara1997} and electric field control \cite{ohno2000} of carrier mediated ferromagnetism. Therefore, it is natural to ask whether the QAH effect can be realized in this system. In this work, we find the Curie temperature of ferromagnetism can be significantly enhanced due to the non-trivial band structure. The quantized Hall conductance appears in a wide regime of parameters below the Curie temperature. Therefore, magnetically doped InAs/GaSb QWs provide an promising platform to search for the QAH effect with a high critical temperature.

\begin{figure}[tb]
    \begin{center}
	    \includegraphics[width=3in,angle=0]{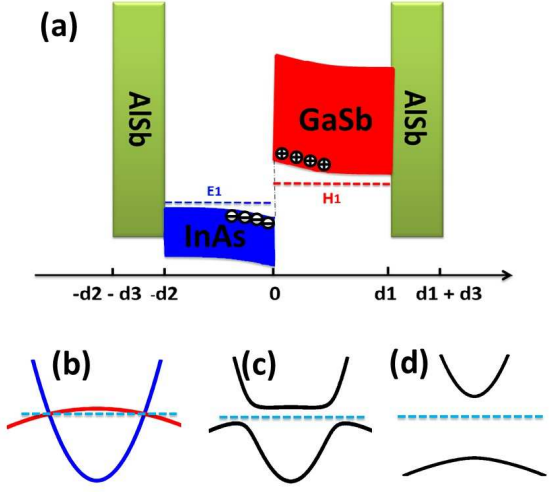} 
    \end{center}
 \caption{
 (Color online) (a) Illustration of an AlSb/InAs/GaSb/AlSb Type-II semiconductor QW. The widths of AlSb, InAs and GaSb are denoted as d$_3$, d$_2$ and d$_1$, respectively. In the inverted regime, hole subbands of GaSb are located above electron subbands of InAs, leading to an electric charge transfer between these two layers. (b), (c) and (d) Band structures for InAs/GaSb QW with $A =0 eV \cdot \AA{}$, $A =0.3 eV \cdot \AA{}$ and $A =2 eV \cdot \AA{}$, respectively; A is the coupling strength between the first hole subband of GaSb and the first electron subbband of InAs.}
    \label{fig:schematic}
\end{figure}

{\it InAs/GaSb quantum wells and ferromagnetism} -
A schematic plot is presented in Fig.\ref{fig:schematic} (a) for InAs/GaSb QWs, in which InAs and GaSb together serve as well layers and AlSb layers are for barriers\cite{Changss1980,Yangprl1997,Chaoprb2000}. The unique feature of InAs/GaSb QWs is that the conduction band minimum of InAs has lower energy than the valence band maximum of GaSb, due to the large band-offset. Consequently, when the well thickness is large enough, the first electron subband of InAs layers, denoted as $\vert E_1 \rangle$, lies below the first hole subband of GaSb layers, denoted as $\vert H_1 \rangle$. This special band alignment is similar to that in HgTe QWs, known as an ``inverted band structure'', which is essential for the QSH effect \cite{Liucprl2008,Knezprl2011}. Since the $\vert H_1 \rangle$ state has a higher energy than the $\vert E_1 \rangle$ state, there is an intrinsic charge transfer between InAs and GaSb layers. This can be seen from the energy dispersion in Fig.\ref{fig:schematic} (b), where the Fermi energy (dashed line) crosses both the $\vert E_1 \rangle$ band in InAs layer (blue line) and the $\vert H_1 \rangle$ band in GaSb layer (red line) if we neglect the coupling between two layers. Clearly, InAs layer is electron doped while GaSb is hole doped. With magnetic doping, free carriers in InAs and GaSb layers are able to mediate exchange coupling between magnetic moments through Ruderman-Kittel-Kasuya-Yosida (RKKY) interaction\cite{Dietlsci2000,Dietlprb2001,Jungwirthrmp2006,Satormp2010}, leading to ferromagnetism in these systems with a Curie temperature $T_c\sim 25K$\cite{NishitaniPE2010}.
Therefore, we expect that magnetically doped InAs/GaSb QWs in the metallic phase of Fig.\ref{fig:schematic} (b) should also be ferromagnetic. The problem is complicated by the coupling between two layers, which induces a hybridization gap, as shown in Fig.\ref{fig:schematic} (c) or (d). Therefore, it is natural to ask what happens to ferromagnetism for the insulating regime with the Fermi energy in the hybridization gap. Below we will answer this question by studying a four band model.

The low energy physics of magnetically doped InAs/GaSb QWs can be well described by a four band model, which was first developed by Bernevig, Hughes and Zhang (BHZ) for HgTe QWs\cite{Bernevigsci2006,Liucprl2008}. In the basis of $\vert E_1 +\rangle$, $\vert H_1 +\rangle$, $\vert E_1-\rangle$ and $\vert H_1 -\rangle$, the effective Hamiltonian can be expressed as
\begin{eqnarray}
	H = H_0 + H_{BIA} + H_{SIA} + H_{ex}.
	\label{eq:Ham}
\end{eqnarray}
The BHZ Hamiltonian $H_0$ is given by
\begin{eqnarray}
  \nonumber&&H_0= \epsilon_k \mathbf{1}_{4 \times 4} +
  \mathcal{M}(\vec{k}) \sigma_0 \otimes \tau_z + Ak_x \sigma_z \otimes \tau_x -Ak_y \sigma_0 \otimes \tau_y
\label{eqn:ham_0}
\end{eqnarray}
where $\epsilon_k = C - D (k^2_x+k^2_y)$, $\mathcal{M}(\vec{k}) = M_0 - B (k^2_x+k^2_y)$ , $\mathbf{1}_{4 \times 4}$ is a 4 by 4 identity matrix, $\sigma$ and $\tau$ are Pauli matrices, representing spin and sub-bands, respectively. All the parameters used below are given in the appendix\cite{Wangappendix2013} for InAs/GaSb QWs. The linear term ($A$ term) couples electron subbands of InAs and hole subbands of GaSb and opens a hybridization gap. $H_{BIA}$ and $H_{SIA}$ describe bulk inversion asymmetry and structural inversion asymmetry\cite{Liucprl2008,Wangappendix2013}. $H_{ex}$ describes the exchange coupling between magnetic moments and electron spin. In the basis of the four band model, we can write $H_{ex}$ as
\begin{eqnarray}
	H_{ex} = \sum_{\vec{R}_n}{\bf S}_M(\vec{R}_n) \cdot\tilde{\bf s},
	\label{eq:Hex}
\end{eqnarray}
where ${\bf S}_M(\vec{R}_n)$ denotes magnetic impurity spin at the position $\vec{R}_n$ and $\tilde{\bf s}$ is regarded as an effective spin operator of the four band model. We should emphasize that both the total angular momentum and exchange coupling constants are included in $\tilde{\bf s}$ for simplicity (See appendix for details\cite{Wangappendix2013}). We only consider magnetization along the z-direction and the operator $\tilde{s}_z$ can be decomposed into
\begin{eqnarray}
	\tilde{s}_z = s_1 \sigma_z \otimes \tau_z + s_2 \sigma_z \otimes \tau_0,
  \label{eqn:spin}
\end{eqnarray}
where $(s_1+s_2)\sigma_z$ ($(-s_1+s_2)\sigma_z$) describes the effective spin operator for conduction (valence) band in the four band model.

Next we determine the Curie temperature of ferromagnetism in this system through the standard mean field theory. With linear response theory, the spin susceptibility for the operator $\tilde{s}_z$ is given by
\begin{eqnarray}
\nonumber&&\tilde{\chi}_s = {\lim_{q \rightarrow 0}}Re[\sum_{i,j,\sigma,\sigma',\vec{k}}\\
    && \frac{\vert \langle u_{i\sigma,\vec{k}} \vert \tilde{s} \vert u_{j\sigma',\vec{k}+\vec{q}} \rangle \vert ^2 (f_{i\sigma}(\vec{k})-f_{j\sigma'}(\vec{k}+\vec{q}))}{E_{j\sigma'}(\vec{k}+\vec{q})-E_{i\sigma}(\vec{k})+ i \Gamma}]
\label{eqn:chi_s}
\end{eqnarray}
where $i,j$ denote conduction and valence bands, $\sigma,\sigma'$ are spin indices, $u_{i\sigma}$ is the eigen wave function with the energy E$_{i\sigma}$, $f_{i\sigma}(\vec{k})$ is the Fermi-Dirac distribution function and $\Gamma$ is band broadening, estimated as $\sim10^{-4}$eV\cite{Knezprb2010}.
The susceptibility of magnetic moment takes the form $\tilde{\chi}_{M} = \frac{S_0(S_0+1)}{3k_BT}$, which is obtained from the dilute limit of Curie-Weiss behavior. $S_0$ is the spin magnitude of magnetic impurities.
The Curie temperature can be determined by the condition $N_0x_{eff}\tilde{\chi}_s(T_c) \tilde{\chi}_M(T_c) = 1$ \cite{Dietlprb2001,Wangappendix2013} in the mean field approximation, where $N_0$ is the cation concentration and $x_{eff}$ is the effective composition of magnetic atoms.

\begin{figure}[tb]
	\includegraphics[height=2.5in,angle=0]{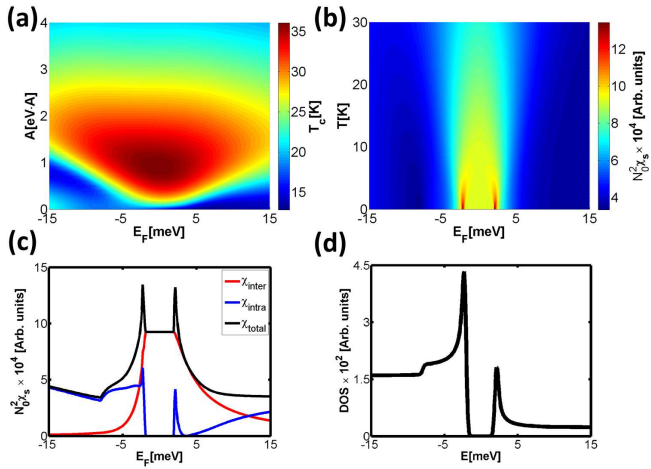}
\caption{
    (Color online). (a) Curie temperatures as a function of the parameter A and Fermi energy. (b) Total spin susceptibility $\chi_s$ as a function of Fermi energy $E_f$ and temperature. (c) Different contribution (intra-band and inter-band contribution) to spin susceptibility $\chi_s$ at $T = 0K$, respectively. (d) Density of states for InAs/GaSb quantum well with a hybridization gap.
  }
    \label{fig:Tc_Ef_A_sus}
\end{figure}

As described above, in the absence of the coupling between two layers, the system must be in a ferromagnetic phase. This corresponds to the case with $A=0$ in the four band model. Therefore, we treat $A$ as a parameter and plot the calculated Curie temperature $T_c$ as a function of $A$ and Fermi energy $E_f$, as shown in Fig. \ref{fig:Tc_Ef_A_sus}(a). When the Fermi energy lies in the hybridization gap (around 0meV), the Curie temperature first increases and then decreases with the increasing of $A$. Therefore, we find surprisingly that the opening of a small hybridization gap will enhance ferromagnetism.

To understand the underlying physics, we consider different origins of spin susceptibility. According to Eq. (\ref{eqn:chi_s}), spin susceptibility can be separated into two parts: the intra-band contribution ($i=j$), and inter-band contribution ($i\neq j$). The intra-band contribution originates from the states near Fermi energy and mediates the RKKY type of coupling between magnetic moments. It is the main origin for ferromagnetism in metallic systems. Indeed, our calculation shows that the intra-band contribution has a maximum around band edge because of the singularity of density of states (Fig. \ref{fig:Tc_Ef_A_sus}(d))\cite{WangjarXiv2012,xu2013}, but is significantly reduced when the Fermi energy is tuned into the band gap (the blue line in Fig. \ref{fig:Tc_Ef_A_sus}(c)). On the other hand, the inter-band contribution mainly originates from the hybridization of wave functions between conduction and valence bands due to the inverted band structure, as discussed in Ref. \onlinecite{Yusci2010}. Therefore, the inter-band contribution shows a peak in the insulating regime and diminishes as the Fermi energy goes away from the band gap. Taking into account both contributions, we find sharp peaks of the total spin susceptibility near band edges at low temperatures (below 4 K), as shown in Fig.\ref{fig:Tc_Ef_A_sus}(b). With increasing temperatures, both peaks are smeared and the spin susceptibility reveals a broad peak structure around band gap, leading to the enhancement of ferromagnetic Curie temperature $T_c$ in the insulating regime.

It is necessary to compare magnetic mechanism in magnetically doped InAs/GaSb QWs with that in Mn doped HgTe QWs and Cr doped (Bi,Sb)$_2$Te$_3$. The low energy effective Hamiltonian of HgTe QWs takes the same form as the Hamiltonian in Eq. \ref{eqn:ham_0}. However, there is one essential difference: the parameter $A$ is much larger in HgTe QWs because electron and hole subbands are in the same layer and coupled strongly. For a large $A$, spin susceptibility will be suppressed due to the large band gap and the disappearance of band edge singularity, as shown in Fig.\ref{fig:schematic} (d).
Consequently, ferromagnetism is not favorable in Mn doped HgTe QWs\cite{Liucprl20082}. The strong interband contribution in our case is similar to that in Cr doped (Bi,Sb)$_2$Te$_3$. The $s_1$ term of the effective spin operator (Eq. (\ref{eqn:spin})) takes the same form as that in the effective model of Cr doped (Bi,Sb)$_2$Te$_3$ (See Ref. \onlinecite{Yusci2010}), which mainly contributes to the inter-band spin susceptibility.
Eq. (\ref{eqn:spin}) includes an additional part $s_2 \sigma_z \otimes \tau_0$, which dominates the intra-band contribution. Our calculation shows that both parts of the spin operator have a significant contribution to spin susceptibility in the case of a small hybridization gap and a band edge singularity. Therefore, in the regime favorable for QAH, a relatively high Curie temperature for ferromagnetism ($T_C \sim 30$ K ) is expected for magnetically doped InAs/GaSb QWs in comparison with the Cr-doped Bi chalcogenides.

{\it Quantized Hall transport and realistic systems} -
Our calculations clearly show that ferromagnetism can be developed in magnetically doped InAs/GaSb QWs. Below $T_c$, magnetic moments align and induce a Zeeman type spin splitting for both conduction and valence bands due to exchange coupling. To realize the QAH states, spin splitting needs to exceed the band gap. This situation is similar to that of Mn-doped HgTe QWs. From Ref. \onlinecite{Liucprl20082}, we find that two conditions for the QAH effect should be satisfied: (1) one spin block becomes a normal band ordering while the other spin block remains in an inverted band ordering;
and (2) the system stays in an insulating state \cite{Wangappendix2013}.
The first condition is satisfied in InAs/GaSb QWs by controlling magnetic doping\cite{Changss1980,Yangprl1997,Chaoprb2000}, while the second condition can be achieved by tuning well thickness. Once these two conditions are satisfied, the QAH effect is expected.

\begin{figure}[tb]
	\includegraphics[width=3.4in,angle=0]{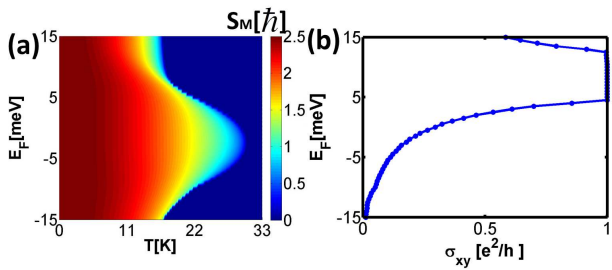}
\caption{(Color online). (a) Magnetization of Mn as a function of temperature and the Fermi energy.  (b) The Hall conductance as a function of Fermi energy $E_F$ at the temperature $T = 1 K$. }
\label{fig:Mag_Ef_T}
\end{figure}

At a low temperature, the average spin $\langle S_M\rangle$ of magnetic atoms and the average effective spin polarization $\langle \tilde{s}_z\rangle$ can be numerically calculated self-consistently\cite{Jungwirthprb1999,Wangappendix2013}. The magnetization of magnetic dopants as a function of the Fermi level and temperature is shown in Fig.\ref{fig:Mag_Ef_T} (a). The critical temperature for ferromagnetic order determined in Fig. \ref{fig:Mag_Ef_T} (a) is consistent with the early calculation based on spin susceptibility. With the obtained magnetization, we compute the Hall conductivity at $T=1K$ with the standard Kubo formula\cite{Thoulessprl1982,Sinitsynprl2006}.  As seen in Fig. \ref{fig:Mag_Ef_T} (b), the Hall conductance is quantized at a value of $e^2/h$ when the Fermi energy falls in band gap and decreases in the metallic regime.

\begin{figure}[tb]
	\includegraphics[width = 0.9\columnwidth,angle=0]{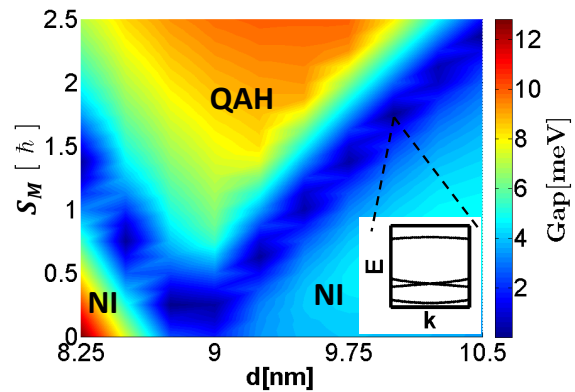}
  \caption{
  (Color online). The band gap is shown as a function of the well thickness of InAs layer and the spin of magnetic impurities. The blue color in the figure shows a gap closing line separating the QAH phase from the normal insulator (NI) phase. The inset image shows a Dirac dispersion at the transition point. All the parameters for Kane model in the calculation are shown in appendix \cite{Wangappendix2013}.
  }
    \label{fig:phase}
\end{figure}

Two key ingredients in the above analysis are the small hybridization gap and band edge singularity, which have been observed in transport experiments of InAs/GaSb QWs \cite{Dularxiv2013,Knezfp2012,Knezprl2011}. Therefore, although our calculation is based on a simple four band model, all the arguments should remain valid qualitatively in realistic materials. Quantitatively, to determine the regime of the QAH effect, we perform an electronic band structure calculation with an eight-band Kane model\cite{li2009,zakharova2001}. The band gap as a function of $d_{InAs}$ and spin of magnetic atom $S_M$ is plotted in Fig.\ref{fig:phase}, from which we can extract the phase diagram. With increasing magnetization, we find a gap closing line in the phase diagram, at which the energy dispersion reveals a single Dirac cone type of band crossing, as shown in the inset of Fig. \ref{fig:phase}. The Hall conductance will change by $\pm e^2/h$ across a Dirac cone type of transition. Therefore, the system is in the QAH phase for large magnetization. With an experimentally achievable magnetic doping concentration\cite{Wangappendix2013}, our calculation gives a band gap as high as 10 meV for the QAH phase. The exchange coupling induced band gap is large enough to host the QAH effect at a high temperature.

{\it Discussion and Conclusion} -
In conclusion, we have proposed a promising material system for the observation of QAH states at relatively high temperatures ($T \sim 30$ K). The materials involved -- Mn-doped InAs/GaSb QWs -- are already well-known to show carrier-mediated ferromagnetism \cite{Ohnoprl1992,AbePE2000,NishitaniPE2010,Dietlnm2010}. In the absence of Mn-doping and at zero bias, InAs/GaSb QWs have electron-type carriers with a concentration of  $\sim 7\times 10^{11} $ cm$^{-2}$\cite{Knezprb2010}. Since Mn-doping adds holes, we estimate a doping of 0.014\% Mn atoms to compensate electron carriers and to shift the chemical potential to the hybridization gap. Additional Mn doping will introduce p-type carriers, which can be diminished by tuning the front and back gates.
To further increase Mn doping, a compensation doping might be required.
Another advantage of InAs/GaSb QWs is the high sample quality with potentially a large mobility of 6,000$cm^2/Vs$ for p-type carriers in non-magnetic heterostructures \cite{Dularxiv2013}, although it is expected to be somewhat smaller in Mn-doped samples \cite{matsuda2004}. Due to the strong exchange coupling, the band gap of the QAH state is able to reach 10 meV, far above the Curie temperature of ferromagnetism (around 30 K). Thus, a well-defined quantized Hall conductance plateau will be expected when the temperature is below Curie temperature. The corresponding experiment is feasible in the present experimental condition. Our calculation based on the standard Zener model has shown a high critical temperature for the QAH effect in magnetically doped InAs/GaSb QWs, which will provide a basis for new spintronics devices with low dissipation.

We would like to thank Kai Chang, Rui-Rui Du, Fu-Chun Zhang, Jian-Hua Zhao and Yi Zhou for useful discussions. This work is supported by the Defense Advanced Research Projects Agency Microsystems Technology Office, MesoDynamic Architecture Program (MESO) through the contract numbers N66001-11-1-4105 and N66001-11-1-4110, and in part by FAME, one of six centers of STARnet, a Semiconductor Research Corporation program sponsored by MARCO and DARPA.

\bibliography{QAH_InAsGaSb}

\begin{thebibliography}{10}%
\makeatletter
\providecommand \@ifxundefined [1]{%
 \ifx #1\undefined \expandafter \@firstoftwo
 \else \expandafter \@secondoftwo
\fi
}%
\providecommand \@ifnum [1]{%
 \ifnum #1\expandafter \@firstoftwo
 \else \expandafter \@secondoftwo
\fi
}%
\providecommand \enquote [1]{``#1''}%
\providecommand \bibnamefont  [1]{#1}%
\providecommand \bibfnamefont [1]{#1}%
\providecommand \citenamefont [1]{#1}%
\providecommand\href[0]{\@sanitize\@href}%
\providecommand\@href[1]{\endgroup\@@startlink{#1}\endgroup\@@href}%
\providecommand\@@href[1]{#1\@@endlink}%
\providecommand \@sanitize [0]{\begingroup\catcode`\&12\catcode`\#12\relax}%
\@ifxundefined \pdfoutput {\@firstoftwo}{%
 \@ifnum{\z@=\pdfoutput}{\@firstoftwo}{\@secondoftwo}%
}{%
 \providecommand\@@startlink[1]{\leavevmode\special{html:<a href="#1">}}%
 \providecommand\@@endlink[0]{\special{html:</a>}}%
}{%
 \providecommand\@@startlink[1]{%
  \leavevmode
  \pdfstartlink
   attr{/Border[0 0 1 ]/H/I/C[0 1 1]}%
   user{/Subtype/Link/A<</Type/Action/S/URI/URI(#1)>>}%
  \relax
 }%
 \providecommand\@@endlink[0]{\pdfendlink}%
}%
\providecommand \url  [0]{\begingroup\@sanitize \@url }%
\providecommand \@url [1]{\endgroup\@href {#1}{\urlprefix}}%
\providecommand \urlprefix [0]{URL }%
\providecommand \Eprint[0]{\href }%
\@ifxundefined \urlstyle {%
  \providecommand \doi [1]{doi:\discretionary{}{}{}#1}%
}{%
  \providecommand \doi [0]{doi:\discretionary{}{}{}\begingroup
  \urlstyle{rm}\Url }%
}%
\providecommand \doibase [0]{http://dx.doi.org/}%
\providecommand \Doi[1]{\href{\doibase#1}}%
\providecommand \bibAnnote [3]{%
  \BibitemShut{#1}%
  \begin{quotation}\noindent
    \textsc{Key:}\ #2\\\textsc{Annotation:}\ #3%
  \end{quotation}%
}%
\providecommand \bibAnnoteFile [2]{%
  \IfFileExists{#2}{\bibAnnote {#1} {#2} {\input{#2}}}{}%
}%
\providecommand \typeout [0]{\immediate \write \m@ne }%
\providecommand \selectlanguage [0]{\@gobble}%
\providecommand \bibinfo [0]{\@secondoftwo}%
\providecommand \bibfield [0]{\@secondoftwo}%
\providecommand \translation [1]{[#1]}%
\providecommand \BibitemOpen[0]{}%
\providecommand \bibitemStop [0]{}%
\providecommand \bibitemNoStop [0]{.\EOS\space}%
\providecommand \EOS [0]{\spacefactor3000\relax}%
\providecommand \BibitemShut [1]{\csname bibitem#1\endcsname}%
\bibitem{Haldaneprl1988}%
  \BibitemOpen
  \bibfield{author}{%
  \bibinfo {author} {\bibfnamefont{F.~D.~M.}\ \bibnamefont{Haldane}},\ }%
  \bibfield{journal}{%
  \bibinfo {journal} {Phys. Rev. Lett.}\ }%
  \textbf{\bibinfo {volume} {61}},\ \bibinfo {pages} {2015} (\bibinfo {year}
  {1988})%
  \bibAnnoteFile{NoStop}{Haldaneprl1988}%
\bibitem{Qiprb2006}%
  \BibitemOpen
  \bibfield{author}{%
  \bibinfo {author} {\bibfnamefont{X.-L.}\ \bibnamefont{Qi}}, \bibinfo {author}
  {\bibfnamefont{Y.-S.}\ \bibnamefont{Wu}},\ and\ \bibinfo {author}
  {\bibfnamefont{S.-C.}\ \bibnamefont{Zhang}},\ }%
  \bibfield{journal}{%
  \Doi{10.1103/PhysRevB.74.085308}{\bibinfo {journal} {Phys. Rev. B}}\ }%
  \textbf{\bibinfo {volume} {74}},\ \bibinfo {pages} {085308} (\bibinfo {year}
  {2006})%
  \bibAnnoteFile{NoStop}{Qiprb2006}%
\bibitem{Qiprb2008}%
  \BibitemOpen
  \bibfield{author}{%
  \bibinfo {author} {\bibfnamefont{X.-L.}\ \bibnamefont{Qi}}, \bibinfo {author}
  {\bibfnamefont{T.~L.}\ \bibnamefont{Hughes}},\ and\ \bibinfo {author}
  {\bibfnamefont{S.-C.}\ \bibnamefont{Zhang}},\ }%
  \bibfield{journal}{%
  \Doi{10.1103/PhysRevB.78.195424}{\bibinfo {journal} {Phys. Rev. B}}\ }%
  \textbf{\bibinfo {volume} {78}},\ \bibinfo {pages} {195424} (\bibinfo {year}
  {2008})%
  \bibAnnoteFile{NoStop}{Qiprb2008}%
\bibitem{Liucprl20082}%
  \BibitemOpen
  \bibfield{author}{%
  \bibinfo {author} {\bibfnamefont{C.-X.}\ \bibnamefont{Liu}}, \bibinfo
  {author} {\bibfnamefont{X.-L.}\ \bibnamefont{Qi}}, \bibinfo {author}
  {\bibfnamefont{X.}~\bibnamefont{Dai}}, \bibinfo {author}
  {\bibfnamefont{Z.}~\bibnamefont{Fang}},\ and\ \bibinfo {author}
  {\bibfnamefont{S.-C.}\ \bibnamefont{Zhang}},\ }%
  \bibfield{journal}{%
  \Doi{10.1103/PhysRevLett.101.146802}{\bibinfo {journal} {Phys. Rev. Lett.}}\
  }%
  \textbf{\bibinfo {volume} {101}},\ \bibinfo {pages} {146802} (\bibinfo {year}
  {2008})%
  \bibAnnoteFile{NoStop}{Liucprl20082}%
\bibitem{Wuprl2008}%
  \BibitemOpen
  \bibfield{author}{%
  \bibinfo {author} {\bibfnamefont{C.}~\bibnamefont{Wu}},\ }%
  \bibfield{journal}{%
  \Doi{10.1103/PhysRevLett.101.186807}{\bibinfo {journal} {Phys. Rev. Lett.}}\
  }%
  \textbf{\bibinfo {volume} {101}},\ \bibinfo {pages} {186807} (\bibinfo {year}
  {2008})%
  \bibAnnoteFile{NoStop}{Wuprl2008}%
\bibitem{Yusci2010}%
  \BibitemOpen
  \bibfield{author}{%
  \bibinfo {author} {\bibfnamefont{R.}~\bibnamefont{Yu}}, \bibinfo {author}
  {\bibfnamefont{W.}~\bibnamefont{Zhang}}, \bibinfo {author}
  {\bibfnamefont{H.-J.}\ \bibnamefont{Zhang}}, \bibinfo {author}
  {\bibfnamefont{S.-C.}\ \bibnamefont{Zhang}}, \bibinfo {author}
  {\bibfnamefont{X.}~\bibnamefont{Dai}},\ and\ \bibinfo {author}
  {\bibfnamefont{Z.}~\bibnamefont{Fang}},\ }%
  \bibfield{journal}{%
  \bibinfo {journal} {Science}\ }%
  \textbf{\bibinfo {volume} {329}},\ \bibinfo {pages} {61} (\bibinfo {year}
  {2010})%
  \bibAnnoteFile{NoStop}{Yusci2010}%
\bibitem{Dingprb2011}%
  \BibitemOpen
  \bibfield{author}{%
  \bibinfo {author} {\bibfnamefont{J.}~\bibnamefont{Ding}}, \bibinfo {author}
  {\bibfnamefont{Z.}~\bibnamefont{Qiao}}, \bibinfo {author}
  {\bibfnamefont{W.}~\bibnamefont{Feng}}, \bibinfo {author}
  {\bibfnamefont{Y.}~\bibnamefont{Yao}},\ and\ \bibinfo {author}
  {\bibfnamefont{Q.}~\bibnamefont{Niu}},\ }%
  \bibfield{journal}{%
  \Doi{10.1103/PhysRevB.84.195444}{\bibinfo {journal} {Phys. Rev. B}}\ }%
  \textbf{\bibinfo {volume} {84}},\ \bibinfo {pages} {195444} (\bibinfo {year}
  {2011})%
  \bibAnnoteFile{NoStop}{Dingprb2011}%
\bibitem{Zhangprl2012}%
  \BibitemOpen
  \bibfield{author}{%
  \bibinfo {author} {\bibfnamefont{H.}~\bibnamefont{Zhang}}, \bibinfo {author}
  {\bibfnamefont{C.}~\bibnamefont{Lazo}}, \bibinfo {author}
  {\bibfnamefont{S.}~\bibnamefont{Bl\"ugel}}, \bibinfo {author}
  {\bibfnamefont{S.}~\bibnamefont{Heinze}},\ and\ \bibinfo {author}
  {\bibfnamefont{Y.}~\bibnamefont{Mokrousov}},\ }%
  \bibfield{journal}{%
  \Doi{10.1103/PhysRevLett.108.056802}{\bibinfo {journal} {Phys. Rev. Lett.}}\
  }%
  \textbf{\bibinfo {volume} {108}},\ \bibinfo {pages} {056802} (\bibinfo {year}
  {2012})%
  \bibAnnoteFile{NoStop}{Zhangprl2012}%
\bibitem{Changsci2013}%
  \BibitemOpen
  \bibfield{author}{%
  \bibinfo {author} {\bibfnamefont{C.-Z.}\ \bibnamefont{Chang}}, \bibinfo
  {author} {\bibfnamefont{J.}~\bibnamefont{Zhang}}, \bibinfo {author}
  {\bibfnamefont{X.}~\bibnamefont{Feng}}, \bibinfo {author}
  {\bibfnamefont{J.}~\bibnamefont{Shen}}, \bibinfo {author}
  {\bibfnamefont{Z.}~\bibnamefont{Zhang}}, \bibinfo {author}
  {\bibfnamefont{M.}~\bibnamefont{Guo}}, \bibinfo {author}
  {\bibfnamefont{K.}~\bibnamefont{Li}}, \bibinfo {author}
  {\bibfnamefont{Y.}~\bibnamefont{Ou}}, \bibinfo {author}
  {\bibfnamefont{P.}~\bibnamefont{Wei}}, \bibinfo {author}
  {\bibfnamefont{L.-L.}\ \bibnamefont{Wang}}, \bibinfo {author}
  {\bibfnamefont{Z.-Q.}\ \bibnamefont{Ji}}, \bibinfo {author}
  {\bibfnamefont{Y.}~\bibnamefont{Feng}}, \bibinfo {author}
  {\bibfnamefont{S.}~\bibnamefont{Ji}}, \bibinfo {author}
  {\bibfnamefont{X.}~\bibnamefont{Chen}}, \bibinfo {author}
  {\bibfnamefont{J.}~\bibnamefont{Jia}}, \bibinfo {author}
  {\bibfnamefont{X.}~\bibnamefont{Dai}}, \bibinfo {author}
  {\bibfnamefont{Z.}~\bibnamefont{Fang}}, \bibinfo {author}
  {\bibfnamefont{S.-C.}\ \bibnamefont{Zhang}}, \bibinfo {author}
  {\bibfnamefont{K.}~\bibnamefont{He}}, \bibinfo {author}
  {\bibfnamefont{Y.}~\bibnamefont{Wang}}, \bibinfo {author}
  {\bibfnamefont{L.}~\bibnamefont{Lu}}, \bibinfo {author}
  {\bibfnamefont{X.-C.}\ \bibnamefont{Ma}},\ and\ \bibinfo {author}
  {\bibfnamefont{Q.-K.}\ \bibnamefont{Xue}},\ }%
  \bibfield{journal}{%
  \bibinfo {journal} {Science}\ }%
  \textbf{\bibinfo {volume} {340}},\ \bibinfo {pages} {167} (\bibinfo {year}
  {2013})%
  \bibAnnoteFile{NoStop}{Changsci2013}%
\bibitem{Wangzfprl2013}%
  \BibitemOpen
  \bibfield{author}{%
  \bibinfo {author} {\bibfnamefont{Z.~F.}\ \bibnamefont{Wang}}, \bibinfo
  {author} {\bibfnamefont{Z.}~\bibnamefont{Liu}},\ and\ \bibinfo {author}
  {\bibfnamefont{F.}~\bibnamefont{Liu}},\ }%
  \bibfield{journal}{%
  \Doi{10.1103/PhysRevLett.110.196801}{\bibinfo {journal} {Phys. Rev. Lett.}}\
  }%
  \textbf{\bibinfo {volume} {110}},\ \bibinfo {pages} {196801} (\bibinfo {year}
  {2013})%
  \bibAnnoteFile{NoStop}{Wangzfprl2013}%
\bibitem{Wangjprl2013}%
  \BibitemOpen
  \bibfield{author}{%
  \bibinfo {author} {\bibfnamefont{J.}~\bibnamefont{Wang}}, \bibinfo {author}
  {\bibfnamefont{B.}~\bibnamefont{Lian}}, \bibinfo {author}
  {\bibfnamefont{H.}~\bibnamefont{Zhang}}, \bibinfo {author}
  {\bibfnamefont{Y.}~\bibnamefont{Xu}},\ and\ \bibinfo {author}
  {\bibfnamefont{S.-C.}\ \bibnamefont{Zhang}},\ }%
  \bibfield{journal}{%
  \Doi{10.1103/PhysRevLett.111.136801}{\bibinfo {journal} {Phys. Rev. Lett.}}\
  }%
  \textbf{\bibinfo {volume} {111}},\ \bibinfo {pages} {136801} (\bibinfo {year}
  {2013})%
  \bibAnnoteFile{NoStop}{Wangjprl2013}%
\bibitem{Zhangscirep2013}%
  \BibitemOpen
  \bibfield{author}{%
  \bibinfo {author} {\bibfnamefont{X.-L.}\ \bibnamefont{Zhang}}, \bibinfo
  {author} {\bibfnamefont{L.-F.}\ \bibnamefont{Liu}},\ and\ \bibinfo {author}
  {\bibfnamefont{W.-M.}\ \bibnamefont{Liu}},\ }%
  \bibfield{journal}{%
  \bibinfo {journal} {Sci. Rep.}}%
   (\bibinfo {year} {2013})%
  \bibAnnoteFile{NoStop}{Zhangscirep2013}%
\bibitem{Klitzingprl1980}%
  \BibitemOpen
  \bibfield{author}{%
  \bibinfo {author} {\bibfnamefont{K.~v.}\ \bibnamefont{Klitzing}}, \bibinfo
  {author} {\bibfnamefont{G.}~\bibnamefont{Dorda}},\ and\ \bibinfo {author}
  {\bibfnamefont{M.}~\bibnamefont{Pepper}},\ }%
  \bibfield{journal}{%
  \Doi{10.1103/PhysRevLett.45.494}{\bibinfo {journal} {Phys. Rev. Lett.}}\ }%
  \textbf{\bibinfo {volume} {45}},\ \bibinfo {pages} {494} (\bibinfo {year}
  {1980})%
  \bibAnnoteFile{NoStop}{Klitzingprl1980}%
\bibitem{Bernevigsci2006}%
  \BibitemOpen
  \bibfield{author}{%
  \bibinfo {author} {\bibfnamefont{B.~A.}\ \bibnamefont{Bernevig}}, \bibinfo
  {author} {\bibfnamefont{T.~L.}\ \bibnamefont{Hughes}},\ and\ \bibinfo
  {author} {\bibfnamefont{S.-C.}\ \bibnamefont{Zhang}},\ }%
  \bibfield{journal}{%
  \bibinfo {journal} {Science}\ }%
  \textbf{\bibinfo {volume} {314}},\ \bibinfo {pages} {1757} (\bibinfo {year}
  {2006})%
  \bibAnnoteFile{NoStop}{Bernevigsci2006}%
\bibitem{Konigsci2007}%
  \BibitemOpen
  \bibfield{author}{%
  \bibinfo {author} {\bibfnamefont{M.}~\bibnamefont{K\"{o}nig}}, \bibinfo
  {author} {\bibfnamefont{S.}~\bibnamefont{Wiedmann}}, \bibinfo {author}
  {\bibfnamefont{C.}~\bibnamefont{Br¨¹ne}}, \bibinfo {author}
  {\bibfnamefont{A.}~\bibnamefont{Roth}}, \bibinfo {author}
  {\bibfnamefont{H.}~\bibnamefont{Buhmann}}, \bibinfo {author}
  {\bibfnamefont{L.~W.}\ \bibnamefont{Molenkamp}}, \bibinfo {author}
  {\bibfnamefont{X.-L.}\ \bibnamefont{Qi}},\ and\ \bibinfo {author}
  {\bibfnamefont{S.-C.}\ \bibnamefont{Zhang}},\ }%
  \bibfield{journal}{%
  \bibinfo {journal} {Science}\ }%
  \textbf{\bibinfo {volume} {318}},\ \bibinfo {pages} {766} (\bibinfo {year}
  {2007})%
  \bibAnnoteFile{NoStop}{Konigsci2007}%
\bibitem{Liuxprl2013}%
  \BibitemOpen
  \bibfield{author}{%
  \bibinfo {author} {\bibfnamefont{X.}~\bibnamefont{Liu}}, \bibinfo {author}
  {\bibfnamefont{H.-C.}\ \bibnamefont{Hsu}},\ and\ \bibinfo {author}
  {\bibfnamefont{C.-X.}\ \bibnamefont{Liu}},\ }%
  \bibfield{journal}{%
  \Doi{10.1103/PhysRevLett.111.086802}{\bibinfo {journal} {Phys. Rev. Lett.}}\
  }%
  \textbf{\bibinfo {volume} {111}},\ \bibinfo {pages} {086802} (\bibinfo
  {month} {Aug}\ \bibinfo {year} {2013})%
  \bibAnnoteFile{NoStop}{Liuxprl2013}%
\bibitem{Hsuprb2013}%
  \BibitemOpen
  \bibfield{author}{%
  \bibinfo {author} {\bibfnamefont{H.-C.}\ \bibnamefont{Hsu}}, \bibinfo
  {author} {\bibfnamefont{X.}~\bibnamefont{Liu}},\ and\ \bibinfo {author}
  {\bibfnamefont{C.-X.}\ \bibnamefont{Liu}},\ }%
  \bibfield{journal}{%
  \Doi{10.1103/PhysRevB.88.085315}{\bibinfo {journal} {Phys. Rev. B}}\ }%
  \textbf{\bibinfo {volume} {88}},\ \bibinfo {pages} {085315} (\bibinfo {year}
  {2013})%
  \bibAnnoteFile{NoStop}{Hsuprb2013}%
\bibitem{Zhangharxiv2013}%
  \BibitemOpen
  \bibfield{author}{%
  \bibinfo {author} {\bibfnamefont{H.}~\bibnamefont{Zhang}}, \bibinfo {author}
  {\bibfnamefont{J.}~\bibnamefont{Wang}}, \bibinfo {author}
  {\bibfnamefont{G.}~\bibnamefont{Xu}}, \bibinfo {author}
  {\bibfnamefont{Y.}~\bibnamefont{Xu}},\ and\ \bibinfo {author}
  {\bibfnamefont{S.-C.}\ \bibnamefont{Zhang}},\ }%
  \bibfield{journal}{%
  \bibinfo {journal} {arXiv preprint arXiv:1308.0349}}%
   (\bibinfo {year} {2013})%
  \bibAnnoteFile{NoStop}{Zhangharxiv2013}%
\bibitem{Liucprl2008}%
  \BibitemOpen
  \bibfield{author}{%
  \bibinfo {author} {\bibfnamefont{C.}~\bibnamefont{Liu}}, \bibinfo {author}
  {\bibfnamefont{T.~L.}\ \bibnamefont{Hughes}}, \bibinfo {author}
  {\bibfnamefont{X.-L.}\ \bibnamefont{Qi}}, \bibinfo {author}
  {\bibfnamefont{K.}~\bibnamefont{Wang}},\ and\ \bibinfo {author}
  {\bibfnamefont{S.-C.}\ \bibnamefont{Zhang}},\ }%
  \bibfield{journal}{%
  \Doi{10.1103/PhysRevLett.100.236601}{\bibinfo {journal} {Phys. Rev. Lett.}}\
  }%
  \textbf{\bibinfo {volume} {100}},\ \bibinfo {pages} {236601} (\bibinfo {year}
  {2008})%
  \bibAnnoteFile{NoStop}{Liucprl2008}%
\bibitem{Knezprb2010}%
  \BibitemOpen
  \bibfield{author}{%
  \bibinfo {author} {\bibfnamefont{I.}~\bibnamefont{Knez}}, \bibinfo {author}
  {\bibfnamefont{R.~R.}\ \bibnamefont{Du}},\ and\ \bibinfo {author}
  {\bibfnamefont{G.}~\bibnamefont{Sullivan}},\ }%
  \bibfield{journal}{%
  \Doi{10.1103/PhysRevB.81.201301}{\bibinfo {journal} {Phys. Rev. B}}\ }%
  \textbf{\bibinfo {volume} {81}},\ \bibinfo {pages} {201301} (\bibinfo {year}
  {2010})%
  \bibAnnoteFile{NoStop}{Knezprb2010}%
\bibitem{Knezprl2011}%
  \BibitemOpen
  \bibfield{author}{%
  \bibinfo {author} {\bibfnamefont{I.}~\bibnamefont{Knez}}, \bibinfo {author}
  {\bibfnamefont{R.-R.}\ \bibnamefont{Du}},\ and\ \bibinfo {author}
  {\bibfnamefont{G.}~\bibnamefont{Sullivan}},\ }%
  \bibfield{journal}{%
  \Doi{10.1103/PhysRevLett.107.136603}{\bibinfo {journal} {Phys. Rev. Lett.}}\
  }%
  \textbf{\bibinfo {volume} {107}},\ \bibinfo {pages} {136603} (\bibinfo {year}
  {2011})%
  \bibAnnoteFile{NoStop}{Knezprl2011}%
\bibitem{Suzukiprb2013}%
  \BibitemOpen
  \bibfield{author}{%
  \bibinfo {author} {\bibfnamefont{K.}~\bibnamefont{Suzuki}}, \bibinfo {author}
  {\bibfnamefont{Y.}~\bibnamefont{Harada}}, \bibinfo {author}
  {\bibfnamefont{K.}~\bibnamefont{Onomitsu}},\ and\ \bibinfo {author}
  {\bibfnamefont{K.}~\bibnamefont{Muraki}},\ }%
  \bibfield{journal}{%
  \Doi{10.1103/PhysRevB.87.235311}{\bibinfo {journal} {Phys. Rev. B}}\ }%
  \textbf{\bibinfo {volume} {87}},\ \bibinfo {pages} {235311} (\bibinfo {year}
  {2013})%
  \bibAnnoteFile{NoStop}{Suzukiprb2013}%
\bibitem{Dularxiv2013}%
  \BibitemOpen
  \bibfield{author}{%
  \bibinfo {author} {\bibfnamefont{L.}~\bibnamefont{Du}}, \bibinfo {author}
  {\bibfnamefont{I.}~\bibnamefont{Knez}}, \bibinfo {author}
  {\bibfnamefont{G.}~\bibnamefont{Sullivan}},\ and\ \bibinfo {author}
  {\bibfnamefont{R.-R.}\ \bibnamefont{Du}},\ }%
  \bibfield{journal}{%
  \bibinfo {journal} {arXiv preprint arXiv:1306.1925}}%
   (\bibinfo {year} {2013})%
  \bibAnnoteFile{NoStop}{Dularxiv2013}%
\bibitem{von1991}%
  \BibitemOpen
  \bibfield{author}{%
  \bibinfo {author} {\bibfnamefont{S.}~\bibnamefont{Von~Molnar}}, \bibinfo
  {author} {\bibfnamefont{H.}~\bibnamefont{Munekata}}, \bibinfo {author}
  {\bibfnamefont{H.}~\bibnamefont{Ohno}},\ and\ \bibinfo {author}
  {\bibfnamefont{L.}~\bibnamefont{Chang}},\ }%
  \bibfield{journal}{%
  \bibinfo {journal} {Journal of Magnetism and Magnetic Materials}\ }%
  \textbf{\bibinfo {volume} {93}},\ \bibinfo {pages} {356} (\bibinfo {year}
  {1991})%
  \bibAnnoteFile{NoStop}{von1991}%
\bibitem{Ohnoprl1992}%
  \BibitemOpen
  \bibfield{author}{%
  \bibinfo {author} {\bibfnamefont{H.}~\bibnamefont{Ohno}}, \bibinfo {author}
  {\bibfnamefont{H.}~\bibnamefont{Munekata}}, \bibinfo {author}
  {\bibfnamefont{T.}~\bibnamefont{Penney}}, \bibinfo {author}
  {\bibfnamefont{S.}~\bibnamefont{von Moln\'ar}},\ and\ \bibinfo {author}
  {\bibfnamefont{L.~L.}\ \bibnamefont{Chang}},\ }%
  \bibfield{journal}{%
  \Doi{10.1103/PhysRevLett.68.2664}{\bibinfo {journal} {Phys. Rev. Lett.}}\ }%
  \textbf{\bibinfo {volume} {68}},\ \bibinfo {pages} {2664} (\bibinfo {year}
  {1992})%
  \bibAnnoteFile{NoStop}{Ohnoprl1992}%
\bibitem{NishitaniPE2010}%
  \BibitemOpen
  \bibfield{author}{%
  \bibinfo {author} {\bibfnamefont{Y.}~\bibnamefont{Nishitani}}, \bibinfo
  {author} {\bibfnamefont{M.}~\bibnamefont{Endo}}, \bibinfo {author}
  {\bibfnamefont{F.}~\bibnamefont{Matsukura}},\ and\ \bibinfo {author}
  {\bibfnamefont{H.}~\bibnamefont{Ohno}},\ }%
  \bibfield{journal}{%
  \Doi{http://dx.doi.org/10.1016/j.physe.2009.12.054}{\bibinfo {journal}
  {Physica E: Low-dimensional Systems and Nanostructures}}\ }%
  \textbf{\bibinfo {volume} {42}},\ \bibinfo {pages} {2681 } (\bibinfo {year}
  {2010})%
  \bibAnnoteFile{NoStop}{NishitaniPE2010}%
\bibitem{munekata1993}%
  \BibitemOpen
  \bibfield{author}{%
  \bibinfo {author} {\bibfnamefont{H.}~\bibnamefont{Munekata}}, \bibinfo
  {author} {\bibfnamefont{A.}~\bibnamefont{Zaslavsky}}, \bibinfo {author}
  {\bibfnamefont{P.}~\bibnamefont{Fumagalli}},\ and\ \bibinfo {author}
  {\bibfnamefont{R.}~\bibnamefont{Gambino}},\ }%
  \bibfield{journal}{%
  \bibinfo {journal} {Applied physics letters}\ }%
  \textbf{\bibinfo {volume} {63}},\ \bibinfo {pages} {2929} (\bibinfo {year}
  {1993})%
  \bibAnnoteFile{NoStop}{munekata1993}%
\bibitem{samarth2012}%
  \BibitemOpen
  \bibfield{author}{%
  \bibinfo {author} {\bibfnamefont{N.}~\bibnamefont{Samarth}},\ }%
  \bibfield{journal}{%
  \bibinfo {journal} {Nature Materials}\ }%
  \textbf{\bibinfo {volume} {11}},\ \bibinfo {pages} {360} (\bibinfo {year}
  {2012})%
  \bibAnnoteFile{NoStop}{samarth2012}%
\bibitem{chapler2012}%
  \BibitemOpen
  \bibfield{author}{%
  \bibinfo {author} {\bibfnamefont{B.}~\bibnamefont{Chapler}}, \bibinfo
  {author} {\bibfnamefont{S.}~\bibnamefont{Mack}}, \bibinfo {author}
  {\bibfnamefont{L.}~\bibnamefont{Ju}}, \bibinfo {author}
  {\bibfnamefont{T.}~\bibnamefont{Elson}}, \bibinfo {author}
  {\bibfnamefont{B.}~\bibnamefont{Boudouris}}, \bibinfo {author}
  {\bibfnamefont{E.}~\bibnamefont{Namdas}}, \bibinfo {author}
  {\bibfnamefont{J.}~\bibnamefont{Yuen}}, \bibinfo {author}
  {\bibfnamefont{A.}~\bibnamefont{Heeger}}, \bibinfo {author}
  {\bibfnamefont{N.}~\bibnamefont{Samarth}}, \bibinfo {author}
  {\bibfnamefont{M.}~\bibnamefont{Di~Ventra}}, \emph{et~al.},\ }%
  \bibfield{journal}{%
  \bibinfo {journal} {Physical Review B}\ }%
  \textbf{\bibinfo {volume} {86}},\ \bibinfo {pages} {165302} (\bibinfo {year}
  {2012})%
  \bibAnnoteFile{NoStop}{chapler2012}%
\bibitem{fujii2013}%
  \BibitemOpen
  \bibfield{author}{%
  \bibinfo {author} {\bibfnamefont{J.}~\bibnamefont{Fujii}}, \bibinfo {author}
  {\bibfnamefont{B.~R.}\ \bibnamefont{Salles}}, \bibinfo {author}
  {\bibfnamefont{M.}~\bibnamefont{Sperl}}, \bibinfo {author}
  {\bibfnamefont{S.}~\bibnamefont{Ueda}}, \bibinfo {author}
  {\bibfnamefont{M.}~\bibnamefont{Kobata}}, \bibinfo {author}
  {\bibfnamefont{K.}~\bibnamefont{Kobayashi}}, \bibinfo {author}
  {\bibfnamefont{Y.}~\bibnamefont{Yamashita}}, \bibinfo {author}
  {\bibfnamefont{P.}~\bibnamefont{Torelli}}, \bibinfo {author}
  {\bibfnamefont{M.}~\bibnamefont{Utz}}, \bibinfo {author}
  {\bibfnamefont{C.~S.}\ \bibnamefont{Fadley}}, \bibinfo {author}
  {\bibfnamefont{A.~X.}\ \bibnamefont{Gray}}, \bibinfo {author}
  {\bibfnamefont{J.}~\bibnamefont{Braun}}, \bibinfo {author}
  {\bibfnamefont{H.}~\bibnamefont{Ebert}}, \bibinfo {author}
  {\bibfnamefont{I.}~\bibnamefont{Di~Marco}}, \bibinfo {author}
  {\bibfnamefont{O.}~\bibnamefont{Eriksson}}, \bibinfo {author}
  {\bibfnamefont{P.}~\bibnamefont{Thunstr\"om}}, \bibinfo {author}
  {\bibfnamefont{G.~H.}\ \bibnamefont{Fecher}}, \bibinfo {author}
  {\bibfnamefont{H.}~\bibnamefont{Stryhanyuk}}, \bibinfo {author}
  {\bibfnamefont{E.}~\bibnamefont{Ikenaga}}, \bibinfo {author}
  {\bibfnamefont{J.}~\bibnamefont{Min\'ar}}, \bibinfo {author}
  {\bibfnamefont{C.~H.}\ \bibnamefont{Back}}, \bibinfo {author}
  {\bibfnamefont{G.}~\bibnamefont{van~der Laan}},\ and\ \bibinfo {author}
  {\bibfnamefont{G.}~\bibnamefont{Panaccione}},\ }%
  \bibfield{journal}{%
  \bibinfo {journal} {Phys. Rev. Lett.}\ }%
  \textbf{\bibinfo {volume} {111}},\ \bibinfo {pages} {097201} (\bibinfo
  {month} {Aug}\ \bibinfo {year} {2013})%
  \bibAnnoteFile{NoStop}{fujii2013}%
\bibitem{Dietlsci2000}%
  \BibitemOpen
  \bibfield{author}{%
  \bibinfo {author} {\bibfnamefont{T.}~\bibnamefont{Dietl}}, \bibinfo {author}
  {\bibfnamefont{H.}~\bibnamefont{Ohno}}, \bibinfo {author}
  {\bibfnamefont{F.}~\bibnamefont{Matsukura}}, \bibinfo {author}
  {\bibfnamefont{J.}~\bibnamefont{Cibert}},\ and\ \bibinfo {author}
  {\bibfnamefont{D.}~\bibnamefont{Ferrand}},\ }%
  \bibfield{journal}{%
  \bibinfo {journal} {Science}\ }%
  \textbf{\bibinfo {volume} {287}},\ \bibinfo {pages} {1019} (\bibinfo {year}
  {2000})%
  \bibAnnoteFile{NoStop}{Dietlsci2000}%
\bibitem{macdonald2005}%
  \BibitemOpen
  \bibfield{author}{%
  \bibinfo {author} {\bibfnamefont{A.}~\bibnamefont{MacDonald}}, \bibinfo
  {author} {\bibfnamefont{P.}~\bibnamefont{Schiffer}},\ and\ \bibinfo {author}
  {\bibfnamefont{N.}~\bibnamefont{Samarth}},\ }%
  \bibfield{journal}{%
  \bibinfo {journal} {Nature Materials}\ }%
  \textbf{\bibinfo {volume} {4}},\ \bibinfo {pages} {195} (\bibinfo {year}
  {2005})%
  \bibAnnoteFile{NoStop}{macdonald2005}%
\bibitem{Jungwirthrmp2006}%
  \BibitemOpen
  \bibfield{author}{%
  \bibinfo {author} {\bibfnamefont{T.}~\bibnamefont{Jungwirth}}, \bibinfo
  {author} {\bibfnamefont{J.}~\bibnamefont{Sinova}}, \bibinfo {author}
  {\bibfnamefont{J.}~\bibnamefont{Ma\ifmmode~\check{s}\else \v{s}\fi{}ek}},
  \bibinfo {author} {\bibfnamefont{J.}~\bibnamefont{Ku\ifmmode~\check{c}\else
  \v{c}\fi{}era}},\ and\ \bibinfo {author} {\bibfnamefont{A.~H.}\
  \bibnamefont{MacDonald}},\ }%
  \bibfield{journal}{%
  \Doi{10.1103/RevModPhys.78.809}{\bibinfo {journal} {Rev. Mod. Phys.}}\ }%
  \textbf{\bibinfo {volume} {78}},\ \bibinfo {pages} {809} (\bibinfo {year}
  {2006})%
  \bibAnnoteFile{NoStop}{Jungwirthrmp2006}%
\bibitem{Dietlnm2010}%
  \BibitemOpen
  \bibfield{author}{%
  \bibinfo {author} {\bibfnamefont{T.}~\bibnamefont{Dietl}},\ }%
  \bibfield{journal}{%
  \Doi{10.1038/nmat2898}{\bibinfo {journal} {Nature materials}}\ }%
  \textbf{\bibinfo {volume} {9}},\ \bibinfo {pages} {965} (\bibinfo {year}
  {2010})%
  \bibAnnoteFile{NoStop}{Dietlnm2010}%
\bibitem{koshihara1997}%
  \BibitemOpen
  \bibfield{author}{%
  \bibinfo {author} {\bibfnamefont{S.}~\bibnamefont{Koshihara}}, \bibinfo
  {author} {\bibfnamefont{A.}~\bibnamefont{Oiwa}}, \bibinfo {author}
  {\bibfnamefont{M.}~\bibnamefont{Hirasawa}}, \bibinfo {author}
  {\bibfnamefont{S.}~\bibnamefont{Katsumoto}}, \bibinfo {author}
  {\bibfnamefont{Y.}~\bibnamefont{Iye}}, \bibinfo {author}
  {\bibfnamefont{C.}~\bibnamefont{Urano}}, \bibinfo {author}
  {\bibfnamefont{H.}~\bibnamefont{Takagi}},\ and\ \bibinfo {author}
  {\bibfnamefont{H.}~\bibnamefont{Munekata}},\ }%
  \bibfield{journal}{%
  \Doi{10.1103/PhysRevLett.78.4617}{\bibinfo {journal} {Phys. Rev. Lett.}}\ }%
  \textbf{\bibinfo {volume} {78}},\ \bibinfo {pages} {4617} (\bibinfo {year}
  {1997})%
  \bibAnnoteFile{NoStop}{koshihara1997}%
\bibitem{ohno2000}%
  \BibitemOpen
  \bibfield{author}{%
  \bibinfo {author} {\bibfnamefont{H.}~\bibnamefont{Ohno}}, \bibinfo {author}
  {\bibfnamefont{D.}~\bibnamefont{Chiba}}, \bibinfo {author}
  {\bibfnamefont{F.}~\bibnamefont{Matsukura}}, \bibinfo {author}
  {\bibfnamefont{T.}~\bibnamefont{Omiya}}, \bibinfo {author}
  {\bibfnamefont{E.}~\bibnamefont{Abe}}, \bibinfo {author}
  {\bibfnamefont{T.}~\bibnamefont{Dietl}}, \bibinfo {author}
  {\bibfnamefont{Y.}~\bibnamefont{Ohno}},\ and\ \bibinfo {author}
  {\bibfnamefont{K.}~\bibnamefont{Ohtani}},\ }%
  \bibfield{journal}{%
  \bibinfo {journal} {Nature}\ }%
  \textbf{\bibinfo {volume} {408}},\ \bibinfo {pages} {944} (\bibinfo {year}
  {2000})%
  \bibAnnoteFile{NoStop}{ohno2000}%
\bibitem{Changss1980}%
  \BibitemOpen
  \bibfield{author}{%
  \bibinfo {author} {\bibfnamefont{L.}~\bibnamefont{Chang}}\ and\ \bibinfo
  {author} {\bibfnamefont{L.}~\bibnamefont{Esaki}},\ }%
  \bibfield{journal}{%
  \bibinfo {journal} {Surface Science}\ }%
  \textbf{\bibinfo {volume} {98}},\ \bibinfo {pages} {70 } (\bibinfo {year}
  {1980})%
  \bibAnnoteFile{NoStop}{Changss1980}%
\bibitem{Yangprl1997}%
  \BibitemOpen
  \bibfield{author}{%
  \bibinfo {author} {\bibfnamefont{M.~J.}\ \bibnamefont{Yang}}, \bibinfo
  {author} {\bibfnamefont{C.~H.}\ \bibnamefont{Yang}}, \bibinfo {author}
  {\bibfnamefont{B.~R.}\ \bibnamefont{Bennett}},\ and\ \bibinfo {author}
  {\bibfnamefont{B.~V.}\ \bibnamefont{Shanabrook}},\ }%
  \bibfield{journal}{%
  \bibinfo {journal} {Phys. Rev. Lett.}\ }%
  \textbf{\bibinfo {volume} {78}},\ \bibinfo {pages} {4613} (\bibinfo {month}
  {Jun}\ \bibinfo {year} {1997})%
  \bibAnnoteFile{NoStop}{Yangprl1997}%
\bibitem{Chaoprb2000}%
  \BibitemOpen
  \bibfield{author}{%
  \bibinfo {author} {\bibfnamefont{E.}~\bibnamefont{Halvorsen}}, \bibinfo
  {author} {\bibfnamefont{Y.}~\bibnamefont{Galperin}},\ and\ \bibinfo {author}
  {\bibfnamefont{K.~A.}\ \bibnamefont{Chao}},\ }%
  \bibfield{journal}{%
  \bibinfo {journal} {Phys. Rev. B}\ }%
  \textbf{\bibinfo {volume} {61}},\ \bibinfo {pages} {16743} (\bibinfo {year}
  {2000})%
  \bibAnnoteFile{NoStop}{Chaoprb2000}%
\bibitem{Dietlprb2001}%
  \BibitemOpen
  \bibfield{author}{%
  \bibinfo {author} {\bibfnamefont{T.}~\bibnamefont{Dietl}}, \bibinfo {author}
  {\bibfnamefont{H.}~\bibnamefont{Ohno}},\ and\ \bibinfo {author}
  {\bibfnamefont{F.}~\bibnamefont{Matsukura}},\ }%
  \bibfield{journal}{%
  \Doi{10.1103/PhysRevB.63.195205}{\bibinfo {journal} {Phys. Rev. B}}\ }%
  \textbf{\bibinfo {volume} {63}},\ \bibinfo {pages} {195205} (\bibinfo {year}
  {2001})%
  \bibAnnoteFile{NoStop}{Dietlprb2001}%
\bibitem{Satormp2010}%
  \BibitemOpen
  \bibfield{author}{%
  \bibinfo {author} {\bibfnamefont{K.}~\bibnamefont{Sato}}, \bibinfo {author}
  {\bibfnamefont{L.}~\bibnamefont{Bergqvist}}, \bibinfo {author}
  {\bibfnamefont{J.}~\bibnamefont{Kudrnovsk\'y}}, \bibinfo {author}
  {\bibfnamefont{P.~H.}\ \bibnamefont{Dederichs}}, \bibinfo {author}
  {\bibfnamefont{O.}~\bibnamefont{Eriksson}}, \bibinfo {author}
  {\bibfnamefont{I.}~\bibnamefont{Turek}}, \bibinfo {author}
  {\bibfnamefont{B.}~\bibnamefont{Sanyal}}, \bibinfo {author}
  {\bibfnamefont{G.}~\bibnamefont{Bouzerar}}, \bibinfo {author}
  {\bibfnamefont{H.}~\bibnamefont{Katayama-Yoshida}}, \bibinfo {author}
  {\bibfnamefont{V.~A.}\ \bibnamefont{Dinh}}, \bibinfo {author}
  {\bibfnamefont{T.}~\bibnamefont{Fukushima}}, \bibinfo {author}
  {\bibfnamefont{H.}~\bibnamefont{Kizaki}},\ and\ \bibinfo {author}
  {\bibfnamefont{R.}~\bibnamefont{Zeller}},\ }%
  \bibfield{journal}{%
  \Doi{10.1103/RevModPhys.82.1633}{\bibinfo {journal} {Rev. Mod. Phys.}}\ }%
  \textbf{\bibinfo {volume} {82}},\ \bibinfo {pages} {1633} (\bibinfo {year}
  {2010})%
  \bibAnnoteFile{NoStop}{Satormp2010}%
\bibitem{Wangappendix2013}%
  \BibitemOpen
  \bibinfo {note} {See Appendix for our derivation of the four band Hamiltonian
  and self-consistent calculation of ferromagnetism.}%
  \bibAnnoteFile{Stop}{Wangappendix2013}%
\bibitem{WangjarXiv2012}%
  \BibitemOpen
  \bibfield{author}{%
  \bibinfo {author} {\bibfnamefont{J.}~\bibnamefont{Wang}}, \bibinfo {author}
  {\bibfnamefont{H.}~\bibnamefont{Mabuchi}},\ and\ \bibinfo {author}
  {\bibfnamefont{X.-L.}\ \bibnamefont{Qi}},\ }%
  \bibfield{journal}{%
  \bibinfo {journal} {arXiv preprint ArXiv:1209.6597}}%
   (\bibinfo {year} {2012})%
  \bibAnnoteFile{NoStop}{WangjarXiv2012}%
\bibitem{xu2013}%
  \BibitemOpen
  \bibfield{author}{%
  \bibinfo {author} {\bibfnamefont{D.-H.}\ \bibnamefont{Xu}}, \bibinfo {author}
  {\bibfnamefont{J.-H.}\ \bibnamefont{Gao}}, \bibinfo {author}
  {\bibfnamefont{C.-X.}\ \bibnamefont{Liu}}, \bibinfo {author}
  {\bibfnamefont{J.-H.}\ \bibnamefont{Sun}}, \bibinfo {author}
  {\bibfnamefont{F.-C.}\ \bibnamefont{Zhang}},\ and\ \bibinfo {author}
  {\bibfnamefont{Y.}~\bibnamefont{Zhou}},\ }%
  \bibfield{journal}{%
  \bibinfo {journal} {arXiv preprint arXiv:1310.4051}}%
   (\bibinfo {year} {2013})%
  \bibAnnoteFile{NoStop}{xu2013}%
\bibitem{Jungwirthprb1999}%
  \BibitemOpen
  \bibfield{author}{%
  \bibinfo {author} {\bibfnamefont{T.}~\bibnamefont{Jungwirth}}, \bibinfo
  {author} {\bibfnamefont{W.~A.}\ \bibnamefont{Atkinson}}, \bibinfo {author}
  {\bibfnamefont{B.~H.}\ \bibnamefont{Lee}},\ and\ \bibinfo {author}
  {\bibfnamefont{A.~H.}\ \bibnamefont{MacDonald}},\ }%
  \bibfield{journal}{%
  \Doi{10.1103/PhysRevB.59.9818}{\bibinfo {journal} {Phys. Rev. B}}\ }%
  \textbf{\bibinfo {volume} {59}},\ \bibinfo {pages} {9818} (\bibinfo {year}
  {1999})%
  \bibAnnoteFile{NoStop}{Jungwirthprb1999}%
\bibitem{Thoulessprl1982}%
  \BibitemOpen
  \bibfield{author}{%
  \bibinfo {author} {\bibfnamefont{D.~J.}\ \bibnamefont{Thouless}}, \bibinfo
  {author} {\bibfnamefont{M.}~\bibnamefont{Kohmoto}}, \bibinfo {author}
  {\bibfnamefont{M.~P.}\ \bibnamefont{Nightingale}},\ and\ \bibinfo {author}
  {\bibfnamefont{M.}~\bibnamefont{den Nijs}},\ }%
  \bibfield{journal}{%
  \Doi{10.1103/PhysRevLett.49.405}{\bibinfo {journal} {Phys. Rev. Lett.}}\ }%
  \textbf{\bibinfo {volume} {49}},\ \bibinfo {pages} {405} (\bibinfo {year}
  {1982})%
  \bibAnnoteFile{NoStop}{Thoulessprl1982}%
\bibitem{Sinitsynprl2006}%
  \BibitemOpen
  \bibfield{author}{%
  \bibinfo {author} {\bibfnamefont{N.~A.}\ \bibnamefont{Sinitsyn}}, \bibinfo
  {author} {\bibfnamefont{J.~E.}\ \bibnamefont{Hill}}, \bibinfo {author}
  {\bibfnamefont{H.}~\bibnamefont{Min}}, \bibinfo {author}
  {\bibfnamefont{J.}~\bibnamefont{Sinova}},\ and\ \bibinfo {author}
  {\bibfnamefont{A.~H.}\ \bibnamefont{MacDonald}},\ }%
  \bibfield{journal}{%
  \Doi{10.1103/PhysRevLett.97.106804}{\bibinfo {journal} {Phys. Rev. Lett.}}\
  }%
  \textbf{\bibinfo {volume} {97}},\ \bibinfo {pages} {106804} (\bibinfo {year}
  {2006})%
  \bibAnnoteFile{NoStop}{Sinitsynprl2006}%
\bibitem{Knezfp2012}%
  \BibitemOpen
  \bibfield{author}{%
  \bibinfo {author} {\bibfnamefont{I.}~\bibnamefont{Knez}}\ and\ \bibinfo
  {author} {\bibfnamefont{R.-R.}\ \bibnamefont{Du}},\ }%
  \bibfield{journal}{%
  \Doi{10.1007/s11467-011-0204-1}{\bibinfo {journal} {Frontiers of Physics}}\
  }%
  \textbf{\bibinfo {volume} {7}},\ \bibinfo {pages} {200} (\bibinfo {year}
  {2012})%
  \bibAnnoteFile{NoStop}{Knezfp2012}%
\bibitem{li2009}%
  \BibitemOpen
  \bibfield{author}{%
  \bibinfo {author} {\bibfnamefont{J.}~\bibnamefont{Li}}, \bibinfo {author}
  {\bibfnamefont{W.}~\bibnamefont{Yang}},\ and\ \bibinfo {author}
  {\bibfnamefont{K.}~\bibnamefont{Chang}},\ }%
  \bibfield{journal}{%
  \bibinfo {journal} {Physical Review B}\ }%
  \textbf{\bibinfo {volume} {80}},\ \bibinfo {pages} {035303} (\bibinfo {year}
  {2009})%
  \bibAnnoteFile{NoStop}{li2009}%
\bibitem{zakharova2001}%
  \BibitemOpen
  \bibfield{author}{%
  \bibinfo {author} {\bibfnamefont{A.}~\bibnamefont{Zakharova}}, \bibinfo
  {author} {\bibfnamefont{S.}~\bibnamefont{Yen}},\ and\ \bibinfo {author}
  {\bibfnamefont{K.}~\bibnamefont{Chao}},\ }%
  \bibfield{journal}{%
  \bibinfo {journal} {Physical Review B}\ }%
  \textbf{\bibinfo {volume} {64}},\ \bibinfo {pages} {235332} (\bibinfo {year}
  {2001})%
  \bibAnnoteFile{NoStop}{zakharova2001}%
\bibitem{AbePE2000}%
  \BibitemOpen
  \bibfield{author}{%
  \bibinfo {author} {\bibfnamefont{E.}~\bibnamefont{Abe}}, \bibinfo {author}
  {\bibfnamefont{F.}~\bibnamefont{Matsukura}}, \bibinfo {author}
  {\bibfnamefont{H.}~\bibnamefont{Yasuda}}, \bibinfo {author}
  {\bibfnamefont{Y.}~\bibnamefont{Ohno}},\ and\ \bibinfo {author}
  {\bibfnamefont{H.}~\bibnamefont{Ohno}},\ }%
  \bibfield{journal}{%
  \Doi{http://dx.doi.org/10.1016/S1386-9477(00)00100-4}{\bibinfo {journal}
  {Physica E: Low-dimensional Systems and Nanostructures}}\ }%
  \textbf{\bibinfo {volume} {7}},\ \bibinfo {pages} {981 } (\bibinfo {year}
  {2000})%
  \bibAnnoteFile{NoStop}{AbePE2000}%
\bibitem{matsuda2004}%
  \BibitemOpen
  \bibfield{author}{%
  \bibinfo {author} {\bibfnamefont{Y.}~\bibnamefont{Matsuda}}, \bibinfo
  {author} {\bibfnamefont{G.}~\bibnamefont{Khodaparast}}, \bibinfo {author}
  {\bibfnamefont{M.}~\bibnamefont{Zudov}}, \bibinfo {author}
  {\bibfnamefont{J.}~\bibnamefont{Kono}}, \bibinfo {author}
  {\bibfnamefont{Y.}~\bibnamefont{Sun}}, \bibinfo {author}
  {\bibfnamefont{F.}~\bibnamefont{Kyrychenko}}, \bibinfo {author}
  {\bibfnamefont{G.}~\bibnamefont{Sanders}}, \bibinfo {author}
  {\bibfnamefont{C.}~\bibnamefont{Stanton}}, \bibinfo {author}
  {\bibfnamefont{N.}~\bibnamefont{Miura}}, \bibinfo {author}
  {\bibfnamefont{S.}~\bibnamefont{Ikeda}}, \emph{et~al.},\ }%
  \bibfield{journal}{%
  \bibinfo {journal} {Physical Review B}\ }%
  \textbf{\bibinfo {volume} {70}},\ \bibinfo {pages} {195211} (\bibinfo {year}
  {2004})%
  \bibAnnoteFile{NoStop}{matsuda2004}%
\bibitem{Burchjmmm2008}%
  \BibitemOpen
  \bibfield{author}{%
  \bibinfo {author} {\bibfnamefont{K.}~\bibnamefont{Burch}}, \bibinfo {author}
  {\bibfnamefont{D.}~\bibnamefont{Awschalom}},\ and\ \bibinfo {author}
  {\bibfnamefont{D.}~\bibnamefont{Basov}},\ }%
  \bibfield{journal}{%
  \Doi{http://dx.doi.org/10.1016/j.jmmm.2008.08.060}{\bibinfo {journal}
  {Journal of Magnetism and Magnetic Materials}}\ }%
  \textbf{\bibinfo {volume} {320}},\ \bibinfo {pages} {3207 } (\bibinfo {year}
  {2008}),\ ISSN \bibinfo {issn} {0304-8853}%
  \bibAnnoteFile{NoStop}{Burchjmmm2008}%
\bibitem{Winkler2003}%
  \BibitemOpen
  \bibfield{author}{%
  \bibinfo {author} {\bibfnamefont{R.}~\bibnamefont{Winkler}},\ }%
  \emph{\bibinfo {title} {Spin-Orbit Coupling Effects in Two-Dimensional
  Electron and Hole Systems}}\ (\bibinfo {publisher} {Springer},\ \bibinfo
  {year} {2003})%
  \bibAnnoteFile{NoStop}{Winkler2003}%
\bibitem{Ashcroftssp}%
  \BibitemOpen
  \bibfield{author}{%
  \bibinfo {author} {\bibfnamefont{N.~W.}\ \bibnamefont{Ashcroft}}\ and\
  \bibinfo {author} {\bibfnamefont{N.~D.}\ \bibnamefont{Mermin}},\ }%
  \emph{\bibinfo {title} {Solid State Physics}}\ (\bibinfo {publisher} {Cengage
  Learning},\ \bibinfo {year} {1976})%
  \bibAnnoteFile{NoStop}{Ashcroftssp}%
\end{thebibliography}%

\newpage
\begin{appendix}

\section{8-band Kane model}
\label{sec:sup}
The 8-band Kane model in the bulk basis
\begin{eqnarray}
\nonumber&&\vert \Gamma^6,1/2 \rangle = \vert S \rangle \vert \uparrow \rangle
\end{eqnarray}
\begin{eqnarray}
\nonumber&&\vert \Gamma^6,-1/2 \rangle = \vert S \rangle \vert \downarrow \rangle
\end{eqnarray}
\begin{eqnarray}
\nonumber&&\vert \Gamma^8,3/2 \rangle = - \frac{1}{\sqrt{2}}\vert  X + iY \rangle \vert \uparrow \rangle
\end{eqnarray}
\begin{eqnarray}
\nonumber&&\vert \Gamma^8,1/2 \rangle =  \frac{1}{\sqrt{6}}(2 \vert Z \rangle \vert \uparrow \rangle - \vert  X + iY \rangle \vert \downarrow \rangle
\end{eqnarray}
\begin{eqnarray}
\nonumber&&\vert \Gamma^8,-1/2 \rangle =  \frac{1}{\sqrt{6}}(2 \vert Z \rangle \vert \downarrow \rangle + \vert  X - iY \rangle \vert \uparrow \rangle
\end{eqnarray}
\begin{eqnarray}
\nonumber&&\vert \Gamma^8,-3/2 \rangle = \frac{1}{\sqrt{2}}\vert  X - iY \rangle \vert \downarrow \rangle
\end{eqnarray}
\begin{eqnarray}
\nonumber&& \vert \Gamma^7,1/2 \rangle = -\frac{1}{\sqrt{3}} (\vert  Z \rangle \vert \uparrow \rangle + \vert X +iY \rangle \vert \downarrow \rangle
\end{eqnarray}
\begin{eqnarray}
\vert \Gamma^7,1/2 \rangle = -\frac{1}{\sqrt{3}} (- \vert  Z \rangle \vert \downarrow \rangle + \vert X - iY \rangle \vert \uparrow \rangle
\label{eqn:basis}
\end{eqnarray}
can be written as
\begin{widetext}
\begin{eqnarray}
  &&H_{Kane}=\left(
	\begin{array}{cccccccc}
          T&0& -\frac{1}{\sqrt{2}}Pk_+& \sqrt{\frac{2}{3}}Pk_z& \frac{1}{\sqrt{6}} Pk_-&0&-\frac{1}{\sqrt{3}}Pk_z&-\frac{1}{\sqrt{3}}Pk_-\\
          0&T&0 &-\frac{1}{\sqrt{6}} Pk_+& \sqrt{\frac{2}{3}}Pk_z& \frac{1}{\sqrt{2}}Pk_- &-\frac{1}{\sqrt{3}}Pk_+ & \frac{1}{\sqrt{3}}Pk_z\\
          -\frac{1}{\sqrt{2}}Pk_-&0&U+V&-\bar{S}_-&R&0&\frac{1}{\sqrt{2}}\bar{S}_-&-\sqrt{2}R\\
          \sqrt{\frac{2}{3}}Pk_z&-\frac{1}{\sqrt{6}} Pk_-&-\bar{S}^{\dag}_-&U-V&C&R&\sqrt{2}V&-\sqrt{\frac{3}{2}}\tilde{S}_-\\
          \frac{1}{\sqrt{6}} Pk_+&\sqrt{\frac{2}{3}}Pk_z&R^{\dag}&C^{\dag}&U-V&\bar{S}^{\dag}_+&-\sqrt{\frac{3}{2}\tilde{S}_+}&-\sqrt{2}V\\
          0&\frac{1}{\sqrt{2}}Pk_+&0&R^{\dag}&\bar{S}_+&U+V&\sqrt{2}R^\dag&\frac{1}{\sqrt{2}}\bar{S}_+\\
          -\frac{1}{\sqrt{3}}Pk_z&-\frac{1}{\sqrt{3}}Pk_-&\frac{1}{\sqrt{2}}\bar{S}^\dag_-&\sqrt{2}V&-\sqrt{\frac{3}{2}}\tilde{S}^\dag_+&\sqrt{2}R&U-\Delta&C\\
          -\frac{1}{\sqrt{3}}Pk_+&\frac{1}{\sqrt{3}}Pk_z&-\sqrt{2}^\dag&-\sqrt{\frac{3}{2}}\tilde{S}^\dag_-&-\sqrt{2}V&\frac{1}{\sqrt{2}}\bar{S}^\dag_+&C^\dag&U-\Delta
	\end{array}
\label{eq:ham_kane}
	\right)
\end{eqnarray}
\end{widetext}
where
\begin{eqnarray}
\nonumber&&T = E_c(z) + \frac{\hbar^2}{2m_0}[(2F+1)k^2_{||}+k_z(2F+1)k_z]
\end{eqnarray}
\begin{eqnarray}
\nonumber&&U = E_v(z) - \frac{\hbar^2}{2m_0}(\gamma_1 k^2_{||} + k_z \gamma_1 k_z)
\end{eqnarray}
\begin{eqnarray}
\nonumber&&V = - \frac{\hbar^2}{2m_0}(\gamma_2 k^2_{||} - 2 k_z \gamma_2 k_z)
\end{eqnarray}
\begin{eqnarray}
\nonumber&&R = - \frac{\hbar^2}{2m_0}(\sqrt{3} \mu k^2_+ - \sqrt{3} \bar{\gamma}k^2_-)
\end{eqnarray}
\begin{eqnarray}
\nonumber&&\bar{S}_{\pm} = - \frac{\hbar^2}{2m_0} \sqrt{3}k_{\pm}(\{\gamma_3,k_z\} + [\kappa,k_z])
\end{eqnarray}
\begin{eqnarray}
\nonumber&&\tilde{S}_\pm = - \frac{\hbar^2}{2m_0} \sqrt{3}k_{\pm}(\{\gamma_3,k_z\} -\frac{1}{3} [\kappa,k_z])
\end{eqnarray}
\begin{eqnarray}
C =  \frac{\hbar^2}{2m_0}k_-[\kappa,k_z]
\end{eqnarray}

Here $\gamma_1$, $\gamma_2$, $\gamma_3$, $\bar{\gamma} = (\gamma_2 + \gamma_3)/2$ and $\mu = (\gamma_3 - \gamma_2)/2$ are parameters depending on materials; $\{$,$\}$ and $[,]$ are commutative and anticommutative operators. The bulk inversion asymmetrical Hamiltonian is expressed as
\begin{widetext}
\begin{eqnarray}
  &&H_{BIA}=\left(
	\begin{array}{cccccccc}
         0&0&0&0&0&0&0&0\\
         0&0&0&0&0&0&0&0\\
         0&0&0&-\frac{1}{2}C_kk_-&C_kk_z&-\frac{\sqrt{3}}{2}C_kk_-&\frac{1}{2\sqrt{2}}C_kk_+&\frac{1}{\sqrt{2}}C_kk_z\\
         0&0&-\frac{1}{2}C_kk_-&0&\frac{\sqrt{3}}{2}C_kk_+&-C_kk_z&0&-\frac{\sqrt{3}}{2\sqrt{2}}C_kk_+\\
         0&0&C_kk_z&-\frac{\sqrt{3}}{2}C_kk_+&0&-\frac{1}{2}C_kk_+&\frac{\sqrt{3}}{2\sqrt{2}}C_kk_-&0\\
         0&0&-\frac{\sqrt{3}}{2}C_kk_+&-C_kk_z&-\frac{1}{2}C_kk_-&0&\frac{1}{\sqrt{2}}C_kk_z&-\frac{1}{2\sqrt{2}}C_kk_-\\
         0&0&\frac{1}{2\sqrt{2}}C_kk_-&0&\frac{\sqrt{3}}{2\sqrt{2}}C_kk_+&\frac{1}{\sqrt{2}}C_kk_z&0&0\\
         0&0&\frac{1}{\sqrt{2}}C_kk_z&-\frac{\sqrt{3}}{2\sqrt{2}}C_kk_-&0&-\frac{1}{2\sqrt{2}}C_kk_+&0&0
	\end{array}
\label{eq:ham_bia8}
	\right)
\end{eqnarray}
\end{widetext}
where $C_k$ depends on materials.

For magnetically doped semiconductors, the sp-d exchange coupling is described by the phenomenological Kondo-like Hamiltonian,
\begin{eqnarray}
 H_{ex} &&= - \sum_{\vec{R}_n} J(r - \vec{R}_n) {\bf S_M}(\vec{R}_n) \cdot {\bf s}_z
\end{eqnarray}
where $J(r - \vec{R}_n)$ is the phenomenological coupling parameters for the magnetic impurity dopant at random site $\vec{R}_n$, ${\bf S_M}(\vec{R_n})$ is the spin of impurity atoms and $\bf{s}_z$ is the electron spin located at $\vec{r}$. The exchange coupling elements in the mean-field approximation is expressed as
\begin{widetext}
\begin{eqnarray}
 \langle \mu \sigma \vert H_{ex} \vert \mu' \sigma' \rangle && = - \sum_{\vec{R}_n} \langle \mu \vert  J(r -\vec{R}_n) \vert \mu' \rangle  {\bf S_M}(\vec{R}_n) \cdot \langle \sigma \vert {\bf s}_z \vert \sigma' \rangle = - \sum_{\vec{R}_n} \delta_{\mu\mu'}\langle \mu \vert J \vert \mu \rangle S_M(\vec{R}_n) \vec{e} \cdot \langle \sigma \vert {\bf s}_z \vert \sigma' \rangle
\end{eqnarray}
\end{widetext}
where $\mu$, $\mu'$ are the band orbital indices, $\sigma$ and $\sigma'$ denote spins, and $\vec{e}$ is the magnetization direction of ${\bf S_M}$.
The exchange parameter $\langle \mu \vert J \vert \mu \rangle$ depends on the symmetry on the band orbital $\vert \mu \rangle$ that could be simplified as $\alpha = \langle S \vert J \vert S \rangle$ and $\beta =\langle X \vert J \vert X \rangle = \langle Y \vert J \vert Y \rangle = \langle Z \vert J \vert Z \rangle$. In this report, we take $N_0 \alpha = 0.5$ eV and $N_0 \beta = -0.98$ eV for both InAs and GaSb layers\cite{Burchjmmm2008} with $N_0$ taking the value of cation concentration.
In the basis as presented in Eq. (\ref{eqn:basis}), the exchange coupling Hamiltonian reads
\begin{widetext}
\begin{eqnarray}
  &&H_{ex}= \sum_{\vec{R}_n} S_M(\vec{R}_n) \left(
	\begin{array}{cccccccc}
       -\hat{e}_z \alpha&-\hat{e}_- \alpha&0&0&0&0&0&0\\
       -\hat{e}_+ \alpha&\hat{e}_z \alpha&0&0&0&0&0&0\\
       0&0&-\hat{e}_z\beta&-\frac{\sqrt{3}}{3}\hat{e}_-\beta&0&0&-\frac{\sqrt{6}}{3}\hat{e}_-\beta&0\\
       0&0&-\frac{\sqrt{3}}{3}\hat{e}_+\beta&-\frac{1}{3}\hat{e}_z\beta&-\frac{2}{3}\hat{e}_-\beta&0&\frac{4}{3\sqrt{2}}\hat{e}_z\beta&-\frac{\sqrt{2}}{3}\hat{e}_-\beta\\
       0&0&0&-\frac{2}{3}\hat{e}_+\beta&\frac{1}{3}\hat{e}_z\beta&-\frac{\sqrt{3}}{3}\hat{e}_-\beta&\frac{\sqrt{2}}{3}\hat{e}_-\beta&\frac{4}{3\sqrt{2}}\hat{e}_z\beta\\
       0&0&0&0&-\frac{\sqrt{3}}{3}\hat{e}_+\beta&\hat{e}_z\beta&0&\frac{\sqrt{6}}{3}\hat{e}_+\beta\\
       0&0&-\frac{\sqrt{6}}{3}\hat{e}_+\beta&\frac{4}{3\sqrt{2}}\hat{e}_z\beta&\frac{\sqrt{2}}{3}\hat{e}_-\beta&0&\frac{1}{3}\hat{e}_z\beta&\frac{1}{3}\hat{e}_-\beta\\
       0&0&0&-\frac{\sqrt{2}}{3}\hat{e}_+\beta&\frac{4}{3\sqrt{2}}\hat{e}_z\beta&\frac{\sqrt{6}}{3}\hat{e}_-\beta&\frac{1}{3}\hat{e}_+\beta&-\frac{1}{3}\hat{e}_z\beta
	\end{array}
\label{eqn:ham_exch8}
	\right)
\end{eqnarray}
\end{widetext}
where $\hat{e}$ denotes the magnetization direction and $\hat{e}_\pm = \hat{e}_x \pm i \hat{e}_y$.

\section{4-band effective Hamiltonian}
\subsection{BIA and SIA Hamiltonian}
The 4-band effective Hamiltonian $H_{eff}$ could be obtained by projecting the 8-band Hamiltonian on the following four subbands: $\vert E_1 +\rangle$, $\vert H_1 +\rangle$, $\vert E_1-\rangle$ and $\vert H_1 -\rangle$, denoted as $\vert A \rangle$, $\vert B \rangle$, $\vert C \rangle$ and $\vert D \rangle$, correspondingly. According to symmetry, one can write the above four bases as
\begin{eqnarray}
\nonumber &&\vert E_1 + \rangle = f_{A,1} \vert \Gamma^6,1/2 \rangle + f_{A,4} \vert \Gamma^8,1/2 \rangle\\
\nonumber &&\vert H_1 + \rangle = f_{B,3} \vert \Gamma^8,3/2 \rangle \\
\nonumber &&\vert E_1 - \rangle = f_{C,2} \vert \Gamma^6,-1/2 \rangle + f_{C,5} \vert \Gamma^8,-1/2 \rangle\\
&&\vert H_1 + \rangle = f_{D,6} \vert \Gamma^8,-3/2 \rangle
\end{eqnarray}.
Since there are two different atoms in one unit cell that breaks the bulk inversion symmetry, we consider the bulk inversion asymmetry (BIA) contribution\cite{Winkler2003}. The BIA Hamiltonian with projection on $\vert E_1 +\rangle$, $\vert H_1 +\rangle$, $\vert E_1-\rangle$ and $\vert H_1 -\rangle$ bands reads
\begin{eqnarray}
	&&H_{BIA}= \left(
	\begin{array}{cccc}
		0&0&\Delta_ek_+&-\Delta_0\\
		0&0&\Delta_0&\Delta_hk_-\\
		\Delta_ek_-&\Delta_0&0&0\\
		-\Delta_0&\Delta_hk_+&0&0
	\end{array}
\label{eqn:ham_bia}
	\right)
\end{eqnarray}
where the parameters $\Delta_e$, $\Delta_{\theta}$, and $\Delta_h$ can be determined by the QW geometry. Because of lack of inversion symmetry along the growth direction of InAs/GaSb QW, we also consider the structural inversion asymmetry (SIA) contribution, which takes form of
\begin{eqnarray}
	&&H_{SIA}=\left(
	\begin{array}{cccc}
		0&0&i\xi_ek_-&0\\
		0&0&0&0\\
		-i\xi^*_ek_+&0&0&0\\
		0&0&0&0
	\end{array}
\label{eqn:ham_sia}
	\right)
\end{eqnarray}
where $\xi_e$ is another parameter that depends on the QW geometry. BHZ model with BIA and SIA corrections confirms the existence of QSH states in InAs/GaSb quantum wells with a set of appropriate parameters.

\subsection{Exchange Hamiltonian}
With projection of 8-band exchange Hamiltonian on $\vert E_1 +\rangle$, $\vert H_1 +\rangle$, $\vert E_1-\rangle$ and $\vert H_1 -\rangle$ bands, we can obtain the effective 4-band exchange Hamiltonian expressed as
\begin{widetext}
\begin{eqnarray}
  &&H_{ex}=\sum_{\vec{R}_n} S_M(\vec{R}_n) \left(
	\begin{array}{cccc}
		\hat{e}_z(\alpha F_1+ \frac{\beta}{3}  F_4)&0&\hat{e}_{-}(\alpha F_1+ \frac{2\beta}{3}  F_4)&0\\
		0&\hat{e}_z\beta&0&0\\
		\hat{e}_{+}(\alpha F_1+ \frac{2\beta}{3}  F_4)&0&-\hat{e}_z(\alpha F_1+ \frac{\beta}{3}  F_4)&0\\
		0&0&0&-\hat{e}_z\beta
	\end{array}
\label{eqn:ham_ex4}
	\right)
\end{eqnarray}
\end{widetext}
where $F_1 = \langle f_{A,1} \vert f_{A,1} \rangle =  \langle f_{C,2} \vert f_{C,2} \rangle = \langle f_{A,1} \vert f_{C,2} \rangle $ and $ F_4 = \langle f_{A,4} \vert f_{A,4} \rangle =  \langle f_{C,5} \vert f_{C,5} \rangle = \langle f_{A,4} \vert f_{C,5} \rangle $. Since we are interested in the QW growth direction ($\hat{e}_z$ direction), we take the effective exchange Hamiltonian as
\begin{widetext}
\begin{eqnarray}
  H_{ex}=sum_{\vec{R}_n} S_M(\vec{R}_n) \left(
	\begin{array}{cccc}
		\alpha F_1+ \frac{\beta}{3}F_4&0&0&0\\
		0&\beta&0&0\\
		0&0&-(\alpha F_1+ \frac{\beta}{3}  F_4)&0\\
		0&0&0&-\beta
	\end{array}
\label{eqn:ham_ex4short}
	\right)
\end{eqnarray}
\end{widetext}
One could write Eq. \ref{eqn:ham_ex4short} as $H_{ex} = \sum_{\vec{R}_n}  S_M(\vec{R}_n) \tilde{s}_z $ where $\tilde{s}_z$ is the effective spin operator and can be expressed as
\begin{eqnarray}
  &&\tilde{s}_z=\left(
	\begin{array}{cccc}
		\alpha F_1+ \frac{\beta}{3}F_4&0&0&0\\
		0&\beta&0&0\\
		0&0&-(\alpha F_1+ \frac{\beta}{3}  F_4)&0\\
		0&0&0&-\beta
	\end{array}
\label{eqn:sz_tilde}
	\right)
\end{eqnarray}

\section{Susceptibility and self-consistent calculation}
We will derive effective susceptibilities of carriers and magnetic dopants starting from $H_{ex} = \sum_{\vec{R}_n}  S_M(\vec{R}_n) \tilde{s}_z$. One can treat the average magnetization of magnetic impurities as an effective magnetic field $h_{e}=\sum_{\vec{R}_n}  S_M(\vec{R}_n)=N_{mag}\langle S_M\rangle$ that is felt by electron spin $\tilde{s}_z$. Here $\langle S_M\rangle$ gives the average magnetization of a magnetic impurity and $N_{mag} = N_0x_{eff}$, where $N_0$ is the cation concentration and $x_{eff}$ is the effective composition of magnetic dopants. Thus, the effective spin susceptibility for carriers is defined as $\langle \tilde{s}_z\rangle=\tilde{\chi}_s h_e$, given by
\begin{eqnarray}
\nonumber\tilde{\chi}_s = &&{\lim_{q \rightarrow 0}}Re [\sum_{i,j,\sigma,\sigma',\vec{k}} \\
&&\frac{\vert \langle u_{i\sigma,\vec{k}} \vert \tilde{s}_z \vert u_{j\sigma',\vec{k}+\vec{q}} \rangle \vert ^2 (f_{i\sigma}(\vec{k})-f_{j\sigma'}(\vec{k}+\vec{q}))}{E_{j\sigma'}(\vec{k}+\vec{q})-E_{i\sigma}(\vec{k})+ i \Gamma}]
\label{eqn:sus_seff}
\end{eqnarray}
from the second order perturbation theory\cite{Dietlprb2001,Yusci2010}, where $i,j$ denote conduction and valence bands, $\sigma,\sigma'$ are spin indices, $u_{i\sigma}$ is the wave function corresponding to energy state E$_{i\sigma}$ with spin index $\sigma$, $f_{i\sigma}(\vec{k})$ is the Fermi-Dirac distribution function and $\Gamma$ is band broadening.

Similarly, the average value of $\langle \tilde{s}_z\rangle$ provides an effective magnetic field $H_M=\langle \tilde{s}_z\rangle$ for the spin $S_M$ of magnetic impurities. The corresponding susceptibility of $S_M$ is defined as $\langle S_M\rangle=\tilde{\chi}_M H_M$.
In the diluted limit, $\langle S_M \rangle$ can be expressed in an empirical form: $\langle S_M \rangle = S_0 B_s(\frac{S_0 H_M}{k_BT})$, where $B_s$ denotes the Brilluoin function $B_S(x) = \frac{2S+1}{S}coth(\frac{2S+1}{2S}x)-\frac{1}{2S}coth(\frac{x}{2S}) \approx \frac{S+1}{3S}x-\frac{(S+1)(2S^2+2S+1)}{900S^3}x^3$, $S_0$ is the spin magnitude of magnetic dopants, $k_B$ is the Boltzmann constant and T represents temperature\cite{Ashcroftssp}. At temperatures close to critical temperature, $\langle S_M \rangle \approx \frac{S_0(S_0+1)H_M}{3k_BT}$. Therefore, the susceptibility of diluted distributed magnetic dopants reads
\begin{eqnarray}
\tilde{\chi}_M = \frac{S_0(S_0+1)}{3k_BT}
\label{eqn:sus_lmeff}
\end{eqnarray}
Finally, One can obtain a self-consistent equation for effective susceptibilities of carriers and magnetic dopants as $N_{mag}\tilde{\chi}_s(T_c)\tilde{\chi}_M (T_c)= 1$\cite{Dietlprb2001}. The $T_c$ can be solved self-consistently according to the above equation.

\section{Conditions for quantum anomalous Hall state}
In order to obtain QAH state, one need to bring one spin block into a right band order while keeping the other spin block still in an inverted band order. By writing $H_{ex} = Diag(G_E,G_H,-G_E,-G_H)$, one arrives at $|G_E-G_H| > 2|M_0|$. Another requirement for the realization of QAH state is that the system is still being in an insulating state. The second condition reads $2|G_E|  > |B+D|k^2_c - |A k_c|$, with the intersection momenta $k^2_c = \frac{||2M_0| + |G_E - G_H||}{2|B|}$.
The above two conditions can be written in function of $s_1$ and $s_2$ used in the paper:(1)$N_0x_{eff}|s_1 \langle S_M \rangle| > |M_0|$  and (2) $2N_0x_{eff}\langle S_M \rangle|s_1 + s_2|  > \frac{B+D}{2B} ||2M_0|+N_0x_{eff}|s_1|\langle S_M \rangle| - A \sqrt{||2M_0|+N_0x_{eff}|s_1|\langle S_M \rangle|/2|B|}$.

\section{Parameters}
The parameters in 8-band Kane model calculation are listed in Table \ref{tab1}, which can be found in Ref. \onlinecite{Chaoprb2000}. The band offset between InAs and GaSb layers is about 150 meV. The valence band difference between InAs and AlSb layers is taken as 180 meV.
\begin{widetext}
\begin{center}
\begin{table}[htb]
  \centering
  \begin{minipage}[t]{1.\linewidth}
	  \caption{ The parameters of Kane model for InAs, GaSb and AlSb.  }
\label{tab1}
\hspace{-1cm}
\begin{tabular}
[c]{ccccccccccc}\hline\hline
 &a[$\AA{}$]&$E_g$[eV]&$\Delta_{so}$[eV]&P[$eV \cdot \AA$]& $\gamma_1$&$\gamma_2$&$\gamma_3$&$\kappa$&$C_k$&F\\
InAs&6.058& 0.41& 0.38 & 9.19 &1.62&-0.65&0.27&-0.005&-0.01&-0.005 \\
GaSb&6.082& 0.8128& 0.752 & 9.23 & 2.61 &-0.56&0.67&0.33&-0.23& 0.333 \\
AlSb&6.133& 2.32& 0.75 & 8.43 &1.46&-0.33&0.41&-0.92&-0.23&0.465 \\\hline\hline\hline
\label{para_kane}
\end{tabular}
  \end{minipage}
\end{table}
\end{center}
\end{widetext}

The parameters presented in Table \ref{tab2} are used in the effective four band calculation. The compositions of magnetic dopants for InAs and GaSb layers we used are 6 \% and 1 \%, respectively. The effective composition is taken as 3\%.
\begin{widetext}
\begin{center}
\begin{table}[htb]
  \centering
  \begin{minipage}[t]{1.\linewidth}
	  \caption{ The parameters of the four band model for InAs/GaSb quantum wells.   }
\label{tab2}
\hspace{-1cm}
\begin{tabular}
[c]{cccccccccccc}\hline\hline
A[$eV\cdot\AA{}$]&B[$eV\cdot \AA{}^2$]&C[eV]&D[eV]&$M_0$[eV]& $\Delta_z$[eV]\\
0.3& -40& -$2.97 \times 10 ^{-3}$ & -30 & -$ 5 \times 10^{-3} $ & $2\times 10^{-4}$\\\hline\hline $\Delta_{e}$[$eV\cdot\AA{}$]&$\Delta_{h}$[$eV\cdot\AA{}$]&$\chi_e$[$eV\cdot\AA{}$]&$F_1$&$F_4$&$\Gamma$[eV]\\
$6.6 \times 10^{-4}$&$6 \times 10^{-4}$ & -$8 \times 10 ^{-4}$ & 0.52 & 0.48 & $3 \times 10 ^{-4}$ \\\hline\hline
\label{tb:para_eff}
\end{tabular}
  \end{minipage}
\end{table}
\end{center}
\end{widetext}

\end{appendix}

\end{document}